\documentclass[openacc]{rsproca_new}

\usepackage{graphicx}
\usepackage{dcolumn}
\usepackage{bm}
\usepackage{color}

\usepackage[normalem]{ulem}
\usepackage{comment}

\usepackage{cancel}

\newcommand{\bu}{{\bf u}}
\newcommand{\vtot}{\tilde{\bf v}}

\newcommand{\uss}{u_{ss}}
\newcommand{\qss}{q_{ss}}
\newcommand{\Ass}{A_{ss}}
\newcommand{\dA}{A^\prime_{ss}}
\newcommand{\deps}{\varepsilon^\prime_{ss}}

\newcommand{\fw}{{\bf f}}

\newcommand{\Lop}{
\left. \frac{\partial \dot{q}_0}{\partial q_0} \right|_{\uss,\qss}}
\newcommand{\Lhat}{\Lop}

\newcommand{\Lzero}{L_0}
\newcommand{\Lplus}{L_+}

\newcommand{\myD}{\left(\alpha + \beta^2/Re\right)}

\newcommand{\Fzero}{F_{0}}
\newcommand{\ctwo}{c_2}
\newcommand{\cone}{c_1}

\newcommand{\underA}[1]{
\underbrace{#1}_{\rm advection}}
\newcommand{\underP}[1]{
\underbrace{#1}_{\rm pressure}}
\newcommand{\underD}[1]{
\underbrace{#1}_{\rm dissipation}}
\newcommand{\underR}[1]{
\underbrace{#1}_{\rm Re.~stress}}

\newcommand{\ex}{{\bf e}_x}
\newcommand{\ey}{{\bf e}_y}
\newcommand{\ez}{{\bf e}_z}

\newcommand{\xrot}{{\bar x}}
\newcommand{\zrot}{{\bar z}}
\newcommand{\urot}{{\bar u}}
\newcommand{\wrot}{{\bar w}}
\newcommand{\exrot}{{\bf e}_\xrot}
\newcommand{\ezrot}{{\bf e}_\zrot}

\titlehead{Research}

\begin{document}

\title{Model for transitional turbulence in a planar shear flow}

\author{
S. J. Benavides$^{1,2,3}$ and D. Barkley$^{2}$}

\address{$^{1}$School of Aeronautics and Space Engineering, Universidad Polit\'{e}cnica de Madrid, Madrid, Spain\\
$^{2}$Mathematics Institute, University of Warwick, Coventry CV4 7AL, United Kingdom\\
$^{3}$School of Mathematics and Maxwell Institute for Mathematical Sciences, University of Edinburgh, Edinburgh, UK
}

\subject{Fluid mechanics, Applied mathematics, Mathematical modelling}

\keywords{Turbulence, Transition to turbulence, Wall-bounded flows}

\corres{S. J. Benavides\\
\email{Santiago.Benavides@ed.ac.uk}}

\begin{abstract}
A central obstacle to understanding the route to turbulence in wall-bounded flows is that turbulence initially appears only intermittently within an otherwise laminar background. In the case of pipe flow, models have deepened our understanding of turbulent onset by providing valuable theory to complement experiments and simulations. In planar configurations, the large-scale flows associated with intermittent transitional turbulence are considerably more complex than for pipes, limiting our ability to develop models and provide theoretical analyses for these cases. 

We address this challenge here by deriving from the Navier-Stokes equations a simplified model for transitional turbulence in a planar setting. The Reynolds-averaged and turbulent-kinetic-energy equations are projected onto a minimal set of wall-normal modes and justified model closures are used for the Reynolds stresses and turbulent dissipation and transport. The model reproduces phenomena found at the onset of turbulence in planar shear flows, such as turbulent-laminar patterns (turbulent bands) oriented obliquely to the streamwise direction and large-scale flows associated with both stationary patterns and growing turbulent spots. We demonstrate the model's utility by showing that patterns arise with decreasing Reynolds number via a linear instability of uniform turbulence and by deriving a bound on the pattern angle at onset: $0<|\theta_c|<45^\circ$.
\end{abstract}




\maketitle

\section{Introduction}

\begin{figure}[b]
	\centering
	\includegraphics[width=0.9\textwidth]{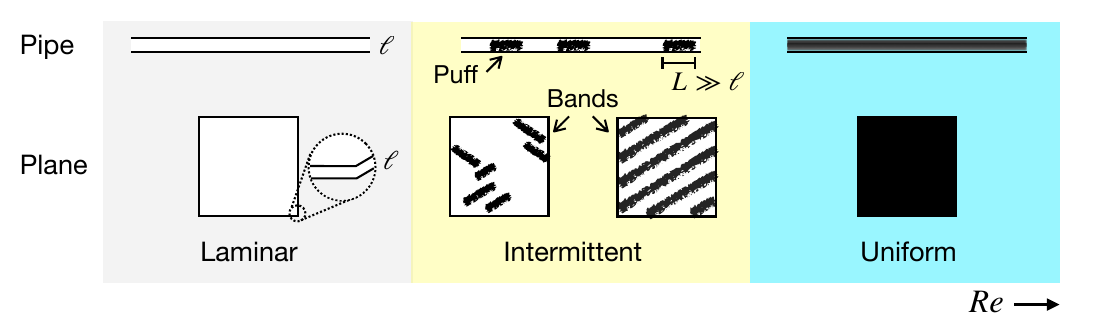}
	\caption{
    Sketch of the basic transition scenario in many wall-bounded shear flows. 
    Flow states are illustrated for a pipe and 
    shear flow between parallel planes, with the inset depicting the geometry rotated into perspective.
    Black signifies turbulent regions, where the fluid motion is highly fluctuating, while white signifies quiescent regions of smooth laminar motion. Three regimes are encountered as a function of the Reynolds number. At low $Re$, laminar motion is the only asymptotic state available to the system. At large $Re$, turbulence can be triggered, and its asymptotic state is spatially uniform, meaning that it adopts the full symmetry afforded by the system. Separating these two extremes is an intermittent regime where, once triggered, turbulence develops into a spatio-temporally intermittent form, composed of turbulent puffs or oblique turbulent bands. 
    Two band states are illustrated: sparse bands (at lower $Re$) and dense bands (at higher $Re$).
    The sketches are not to scale: the length scale $L$ of turbulent structures, whether puffs or bands, is much larger than the wall separation $\ell$, which sets the scale of turbulent fluctuations. Coarse-grained modelling exploits that $\ell/L \ll 1$. 
 } 
    \label{fig:intro_sketch}
\end{figure}

The route to turbulence in many wall-bounded shear flows is mediated by a fascinating regime in which turbulence cannot be sustained throughout the system; rather, it occurs intermittently within laminar flow \cite{manneville2015transition,manneville2016review,manneville2017laminar,tuckerman2020patterns,avila2023transition}. While the laminar state is stable to small perturbations, strongly nonlinear patches of turbulence may be sustained via interactions with neighbouring regions of quiescent laminar flow \cite{wygnanski1973transition,
barkley2007mean, song2017speed, gome2023patterns1}. 
Figure \ref{fig:intro_sketch} illustrates 
the broad characteristics of wall-bounded transition in prototypical geometries: pipes and planar shear flows. 
In pipe flow, the turbulent patches take the form of axially localized turbulent structures known as puffs. Their dynamics and interactions along the pipe are crucial to the intermittent state mediating transition. In planar flows, transition is mediated by turbulent bands, also known as turbulent stripes, tilted obliquely to the streamwise (horizontal) direction. As is clearly evident, transition via turbulent bands is spatially richer than transition via puffs. 

A vast amount is known about both cases, but our theoretical understanding of transition in pipe flow is superior because we are able to reproduce and analyse the transition scenario with simple models. These are systems of partial differential equations that depend only on the axial coordinate and time, based on the interaction between large-scale (coarse-grained) turbulence and mean flow.  
Although the first models based on turbulent mean-flow interaction in pipes appeared more than a decade ago 
\cite{barkley2011simplifying}, there has been no adequate extension of such models to the planar case, and this has hindered our theoretical understanding of the transition to turbulence in wall-bounded flow. We address this here by deriving from the Navier-Stokes equations a simplified model for transitional turbulence in a planar setting.

We expand on the introductory paragraphs, beginning with further details on the intermittent route to turbulence, specifically the boundaries of the intermittent regime. As the flow rate, or Reynolds number ($Re$), decreases, puffs become sparse, and turbulent bands take on a fragmented form until a critical point is reached below which turbulence is no longer sustained. Universal scaling laws associated with directed percolation (DP) have been established for such critical points in some flows~\cite{lemoult2016directed,chantry_universal, klotz2022phase,hof2023directed}.
Although critical phenomena have received considerable attention, this is not the focus of our work. 
As $Re$ increases, laminar regions disappear, and puffs and bands give way to a uniform turbulent state, whose mean recovers the full symmetry of the flow geometry. Equivalently, a symmetry breaking of the uniform turbulent state occurs as $Re$ decreases. A substantial body of work has analyzed and characterized this transition in the planar case \cite{prigent2002large,tuckerman2011patterns, philip2011temporal,manneville2012turing, kashyap2022linear,gome2023patterns1,gome2023patterns2,reetz2019exact}, but the understanding remains incomplete.
This is, in part, the focus of our work.  
Figure~\ref{fig:intro_sketch} is intended to illustrate only generic characteristics of wall-bounded transition. Many variations of such shear flows exist, differing in geometry (e.g., ducts, concentric cylinders, and disks) and in the mechanisms driving the shear (e.g., pressure gradients, moving walls, and body forces). While analogues of puffs and bands are generic, the details of intermittent state and what takes place at the boundaries of the intermittent regime may vary with the particular flow configuration.

In this work we focus on a modelling approach aimed at describing the large-scale (coarse-grained) dynamics of the flows. As illustrated in figure~\ref{fig:intro_sketch}, turbulent structures vary over distances much longer than the constrained direction of shear -- the scale at which turbulence is generated \cite{prigent2002large,prigent2003long, barkley2005computational,duguet2010formation, duguet2013oblique}. Turbulent fluctuations may be viewed as ``microscopic'' \cite{pomeau2015transition, barkley2016theoretical} and averaged over, an approach pioneered by Reynolds \cite{Reynolds1895}.
For pipe flow, models in two scalar fields -- the amplitudes of large-scale turbulence and mean flow -- have proven successful in representing the dynamics of transitional turbulence in a single spatial coordinate (the pipe axis) and time \cite{barkley2011simplifying, barkley2015rise, barkley2016theoretical}. Phenomenological equations capture turbulent puffs as smooth, non-fluctuating structures and provide an underlying bifurcation sequence for the route to turbulence, which can be readily analysed using dynamical systems theory \cite{barkley2016theoretical,frishman2022dynamical}.
The role of ``microscopic'' turbulent fluctuations can be examined separately by incorporating noise terms into the model \cite{pomeau2015transition, barkley2016theoretical}. Other modelling approaches to pipe flow exist. Notable cases are models based on azimuthal zonal flow \cite{shih2016ecological}, which have been extended to include effects of the large-scale streamwise flow \cite{wang2022stochastic}, and models of puffs as discrete objects whose dynamics are modelled from experiments and simulations \cite{lemoult2024directed}.

Turbulent bands are more complex than turbulent puffs, and pipe models do not readily extend to the planar case. As a result, our understanding of bands is less developed than that of puffs. For example, turbulent bands are tilted relative to the streamwise flow direction, with typical angles ranging from approximately $20^\circ$ to $45^\circ$, depending on $Re$ and details of the flow geometry. We lack a fundamental mechanism explaining why bands are tilted and what determines their angle. This is not to say that there is no theoretical understanding of the band angle. 
Numerical studies have imposed, and thereby investigated the range of allowed tilt angles and how this depends on Reynolds number \cite{barkley2007mean,reetz2019exact}.
Furthermore, using a series of numerical simulations, Kashyap {\em et al.} provide evidence of a linear instability of the uniform turbulent state to turbulent bands tilted at $23^\circ \pm 0.5^\circ$ \cite{kashyap2022linear}. The origin of the band angle and the role of linear instability as a mechanism for symmetry breaking are precisely the types of issues that can be effectively addressed through modelling.

The difficulty of extending pipe modelling arises primarily from the large-scale mean flow. In a pipe, the mean flow is effectively unidirectional and can be adequately represented by a single scalar field \cite{barkley2016theoretical}, whereas in the planar case, the mean flow must be described by a vector field (see figure~\ref{fig:geometry}b discussed below). Fluid advection plays a crucial role in these flows \cite{barkley2007mean,duguet2013oblique,gome2023patterns1},
and the vector nature of the mean flow must be properly accounted for if a model is to accurately capture the fluid mechanics of turbulent bands.
There are alternative modelling approaches to planar flows distinct from that pursued in our work. One broad approach is to reduce the wall-normal resolution, either by lowering the resolution in an otherwise standard simulation \cite{manneville2011modelling} or by projecting the Navier–Stokes equations onto a few modal basis functions \cite{chantry2016turbulent,chantry_universal, seshasayanan2015laminar, lagha2007largescaleflow,lagha2007modeling}. Such approaches can capture bands and the vector character of mean flow, but require the simulation of small-scale turbulence fluctuations. Another interesting modelling strategy introduces spatial coupling to the ordinary differential equations describing the self-sustaining process of wall-bounded turbulence \cite{manneville2012turing,kashyap2025laminar}. While these models provide insights into turbulent bands, they lack a vector representation of the large-scale flow and the associated mechanisms.

In this paper, we derive a quantitative, coarse-grained model of planar shear turbulence directly from the Reynolds averaged Navier–Stokes equations through specified truncations and model closures. The large-scale mean flow is represented by the smallest possible set of modes describing a three-dimensional vector field. 
We show that the model successfully captures a wide range of behaviours associated with turbulent bands. The model is readily amenable to analysis, as we demonstrate by identifying a linear instability of uniform turbulence to turbulent bands and deriving a selection criterion for the band angle at the onset of this instability. We leave the issue of fluctuations, and thus the critical phenomena at the boundary between intermittency and laminar flow, to future work.

\section{Model derivation} \label{sec:model_deriv}
\subsection{Geometry and governing equations}\label{subsec:gov_eq}

We derive our model from the most mathematically tractable planar flow that exhibits turbulent-laminar bands: a shear layer between parallel, stress-free boundaries, driven by a sinusoidal body force rather than by wall motion or a pressure gradient.
Studies of this configuration in the inviscid case date back to at least Tollmien \cite{tollmien1935}. In the transition literature, the system is called Waleffe flow (Wf), after Waleffe who used it in his work on the self-sustaining process 
\cite{waleffe1997self}.
It reproduces the same transitional phenomena as plane Couette flow (pCf) and other planar wall-bounded shear flows \cite{chantry2016turbulent,tuckerman2020patterns,chantry_universal}, while the sinusoidal driving force and stress-free boundary conditions significantly simplify both analytical and computational treatments. 

\begin{figure}
	\centering
	\includegraphics[width=0.9\textwidth]{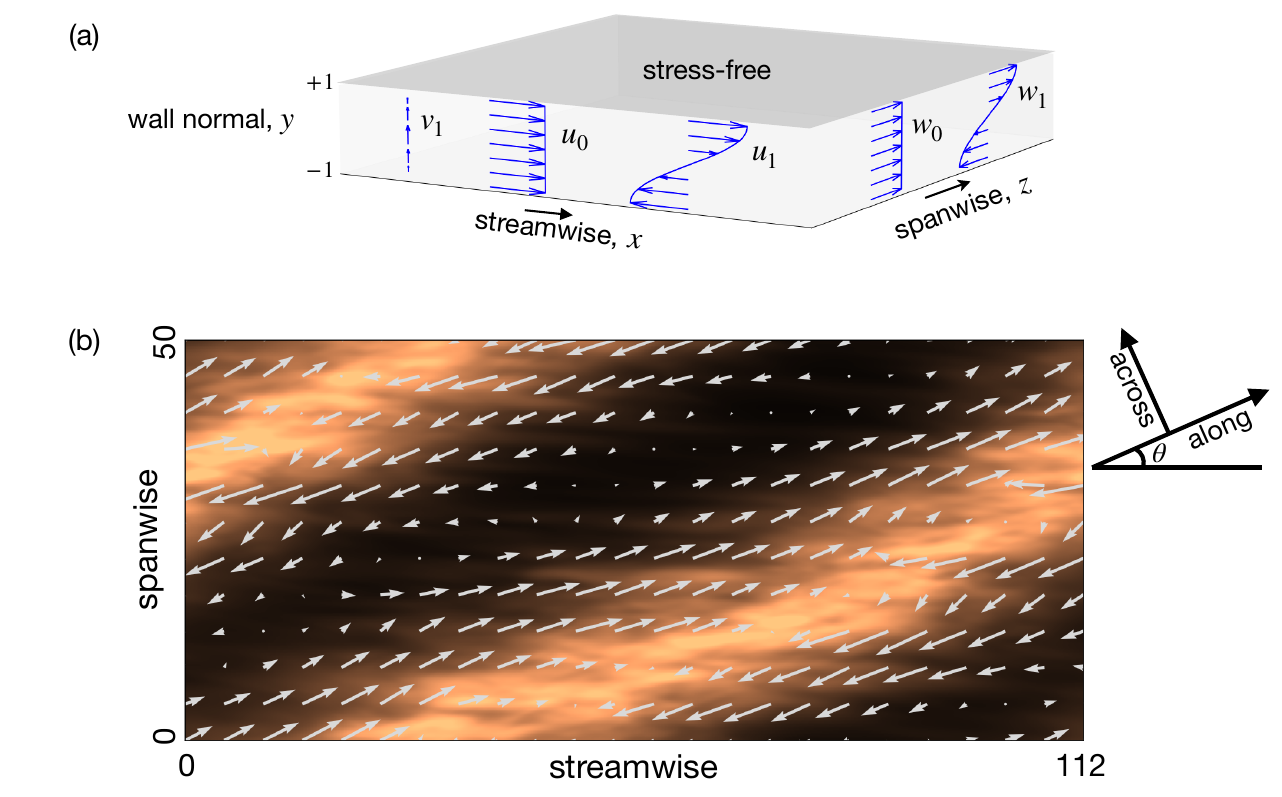}
    \caption{(a)
    Waleffe flow geometry. Fluid is confined between two parallel, stress-free boundaries and is driven by a body force of the form ${\bf f} = f \sin(\pi y /2) {\bf e}_x$. Shown are illustrations of the five wall-normal (vertical) modes in the model used to represent the large-scale flow. 
    The mode labelled $u_1$ has the same form as the forcing and the laminar flow: $\sin( \pi y/2) {\bf e}_x$.
    (b) Turbulent band in Waleffe flow. Visualized is the time- and vertically-averaged turbulent kinetic energy (TKE) (colour) and velocity field (arrows) from a direct numerical simulation at $Re=140$. The averages were taken over 990 advective time units.
    Along- and across-band directions are shown and are referred to frequently in the text. 
    }
    \label{fig:geometry}
\end{figure}

We use coordinates $(x,y,z)$ aligned with the streamwise, wall-normal (vertical), and spanwise directions, respectively. See figure ~\ref{fig:geometry}a. The components of the velocity $\vtot$ are denoted by $(\tilde{u},\tilde{v},\tilde{w})$. We non-dimensionalize using a length scale $h$ equal to the half gap between the stress-free walls, and a velocity scale $U$, equal to the maximum laminar velocity. Thus, time is non-dimensionalized by the advective timescale $h/U$. 
In non-dimensional coordinates, the walls are at $y = \pm 1$. We use $\beta \equiv \pi/2$ for the lowest vertical wavenumber allowed by the geometry.

The governing equations for the flow are the Navier-Stokes equations,
\begin{align}
        \frac{\partial \vtot}{\partial t} + \vtot\cdot \nabla \vtot 
        & = - \nabla \tilde{p} + \frac{1}{Re}\nabla^2 \vtot - \alpha \vtot_H + \fw, \label{eq:vfull} \\
            \nabla \cdot \vtot & = 0,
        \label{eq:div_v}
\end{align}
where $\tilde{p}$ is the pressure, $Re \equiv U h / \nu$ is the Reynolds number, and $\nu$ is the kinematic viscosity. In addition to pressure and viscosity, there are drag and body-force terms on the right-hand side of momentum equation \eqref{eq:vfull}.
The drag term, $-\alpha \vtot_H$, only acts on (and against) the horizontal velocity components $\vtot_H \equiv \tilde{u} {\bf e}_x + \tilde{w} {\bf e}_z$. The parameter $\alpha$ is the drag coefficient and is taken to be small.
Such a drag term, often called Rayleigh or Ekman friction, is used frequently in hydrodynamic modelling contexts to approximate the effect of friction due to a no-slip boundary that has been omitted from the model \cite{suri2014drag,tuckerman2020patterns,chantry_universal,marcus1998model,pedlosky2013geophysical}. It was introduced by Chantry {\em et al.} \cite{chantry_universal} in Wf and is only a significant source of dissipation relative to viscous effects for $y$-independent modes and large horizontal scales. 
Since the goal of the model is to capture flow features on large horizontal scales, Ekman drag plays an important role. As part of our forthcoming model closure, the drag coefficient will ultimately take different values for the $y$-independent and $y$-dependent modes (\S\ref{subsec:params}).

The body force $\fw$ is defined to be,
\begin{equation}
    \fw \equiv \left(\alpha + \frac{\beta^2}{Re}\right) \sin\left(\beta y\right) {\bf e}_x.
\end{equation}
This force drives a simple laminar flow of the form $\vtot_{lam} = \sin(\beta y) {\bf e}_x$. Although the laminar solution has an inflection point, it is nevertheless linearly stable due to wall confinement \cite{waleffe1997self}.

The governing equations are accompanied by periodic boundary conditions in the horizontal $(x,z)$ directions, and stress-free boundary conditions in the vertical direction
\begin{equation}
    \partial_y \tilde{u}(x,\pm 1, z) = \tilde{v}(x,\pm 1, z) = \partial_y \tilde{w}(x,\pm 1, z) = 0,
    \label{eq:BCS_v}
\end{equation}
which includes a no-penetration condition on the wall-normal velocity component. 

To establish the model closures and calibrate the parameter values introduced below, we performed fully-resolved direct numerical simulations (DNS) of the governing equations using the open-source pseudo-spectral code Dedalus \cite{burns2020dedalus}. We simulated the equations in a domain of dimensions $L_x \times 2 \times L_z$, imposing periodic boundary conditions in the horizontal directions.
For time-stepping, we used a Crank-Nicolson method for the linear terms and forward Euler for the nonlinear terms (`CNAB1' solver in Dedalus \cite{burns2020dedalus}), with a time step $\Delta t = 0.003$. A Fourier-spectral method with 3/2 dealiasing was applied in the horizontal directions, with a resolution of eight grid points per spatial unit. In the vertical direction, we used a SinCos basis with the same spatial resolution. We initialized the simulations with random conditions at higher $Re$, and we obtained the banded solution shown in figure~\ref{fig:geometry}b by incrementally decreasing Re from the resulting uniform state.

In addition to the DNS in streamwise–spanwise domains shown in figure~\ref{fig:geometry}b, we performed simulations of turbulent bands using the tilted domain approach introduced by Barkley \& Tuckerman \cite{barkley2005computational}. In this setup, the streamwise direction is shortened, the spanwise direction is extended, and the domain is then tilted at a fixed angle ($24^\circ$ in our case) so that the shortened direction aligns with the along-band direction (figure \ref{fig:geometry}b). 
Averaging such simulations in the along-band and time directions gives profiles depending on the across-band direction only. These are presented in figures throughout this section and are used for final model calibration.

\subsection{Reynolds Average}

We apply Reynolds averaging to the momentum equation by letting $\vtot = \bu + {\bf u^\prime}$, where $\bu \equiv \langle \vtot \rangle$ is the mean velocity field and ${\bf u^\prime}$ is the fluctuating velocity field satisfying $\langle {\bf u^\prime} \rangle = 0$\cite{Reynolds1895}. The components of $\bu$ are denoted by $(u,v,w)$, and similarly $\bu^\prime = (u^\prime,v^\prime,w^\prime)$. We will refer to $\bu$ as the {\em large-scale flow}. 
We define $q \equiv \langle {\bf u^\prime}\cdot{\bf u^\prime} \rangle/2$ and refer to this as the turbulent kinetic energy (TKE), although it should be understood that the fluctuating field ${\bf u^\prime}$ may be very small in some regions, reflecting small fluctuations about the laminar state. 
We are agnostic regarding the specific averaging operation $\langle \cdot \rangle$ used in the Reynolds averaging to defined the large-scale fields. 
In practice, for the flows of interest, ensemble averaging, spatio-temporal averaging, or filtering may be applied, e.g.\ 
\cite{kashyap2022linear, gome2023patterns1, barkley2007mean,Ohnishi_2011,tuckerman2020patterns, duguet2013oblique,wang2020quadrupolar}.
The averaging used below for model calibration is an along-band and time average \cite{barkley2007mean,Ohnishi_2011, tuckerman2020patterns}.

Following standard procedures \cite{pope2000turbulent}, we arrive at the Reynolds-averaged equations of motion for the large-scale flow,
\begin{subequations}
\begin{equation}
    \frac{\partial \bu}{\partial t} + \bu\cdot \nabla \bu = - \nabla p + \frac{1}{Re}\nabla^2 \bu - \alpha \bu_H + \fw + \nabla \cdot {\bf \mathcal{R}}, \label{eq:ufull}
\end{equation}
and TKE,
\begin{equation}
    \frac{\partial q}{\partial t} + \bu \cdot \nabla q + \nabla \cdot \textbf{T} = \mathcal{P} - \varepsilon + \frac{1}{Re} \nabla^2 q - 2\alpha q, \label{eq:qfull}
\end{equation}
\label{eq:full}
\end{subequations}
where $\bu$ satisfies the same incompressibility constraint and boundary conditions as $\vtot$: equations
\eqref{eq:div_v} and \eqref{eq:BCS_v}. The TKE also has a vanishing wall-normal derivative at the two walls. 

A small approximation is already made in equation \eqref{eq:qfull}. The exact drag term from averaging is $\alpha \langle (u^\prime)^2 + (w^\prime)^2 \rangle$, and we have approximated it as $2 \alpha q = \alpha \langle (u^\prime)^2 + (v^\prime)^2 + (w^\prime)^2 \rangle$. Since $\langle (v^\prime)^2 \rangle \ll \langle (u^\prime)^2 + (w^\prime)^2 \rangle$, this approximation has negligible effect. 

The averaging has introduced four new fluctuation correlation terms that will eventually require closure choices. These are the Reynolds stress,
\begin{equation}
    \mathcal{R}_{ij} = -\langle u^\prime_i u^\prime_j \rangle, \label{eq:restress}
\end{equation}
the turbulent production,
\begin{equation}
    \mathcal{P} = -\langle u^\prime_i u^\prime_j \rangle \partial_j u_i,
\end{equation}
the pseudo-dissipation rate,
\begin{equation}
    \varepsilon = \frac{1}{Re}\left\langle \partial_j u_i^\prime \partial_j u_i^\prime \right\rangle,
\end{equation}
and the turbulent transport,
\begin{equation}
        T_i = \frac{1}{2} \langle u^\prime_i u^\prime_j u^\prime_j \rangle + \langle u^\prime_i p^\prime \rangle. \label{eq:transp}
\end{equation}
We use the Einstein summation convention, with $u_i$ representing an indexing of $u,v,w$ and similarly for $u^\prime_i$. Note that closing the Reynolds stress term also closes the production term.

\subsection{Truncation of Vertical Modes}

We turn to modelling.
Our first step is to introduce a Galerkin truncation in the vertical direction. Severe truncations, in both the wall-normal and streamwise directions, have been shown to successfully reproduce turbulent statistics at both low\cite{manneville2011modelling,chantry2016turbulent,chantry_universal, seshasayanan2015laminar, lagha2007largescaleflow,lagha2007modeling} and high\cite{Thomas2014,Constantinou2014,Farrell2017,Cavalieri2022,Minnick2024} Reynolds numbers in a variety of settings such as plane Couette and plane Poiseuille flows. 
Given the sinusoidal body force $\fw$, corresponding sinusoidal laminar flow, and the stress-free boundary conditions, we use a sine-cosine basis and project onto the first two vertical modes (that is, the $y$-independent mode and $\sin(\beta y)$ or $\cos(\beta y)$ mode). The model large-scale flow and TKE are represented by
\begin{subequations}
\begin{eqnarray}
u(x,y,z,t) &=& u_0(x,z,t) + u_1(x,z,t) \sin(\beta y), \label{eq:u_modes} \\
v(x,y,z,t) &=& v_1(x,z,t) \cos(\beta y), \label{eq:v_modes} \\
w(x,y,z,t) &=& w_0(x,z,t) + w_1(x,z,t) \sin(\beta y), \label{eq:w_modes} \\
q(x,y,z,t) &=& q_0(x,z,t) + q_1(x,z,t)\sin(\beta y),
\label{eq:q_modes}
\end{eqnarray}
\label{eq:modes}
\end{subequations}
where indices now represent vertical mode number, rather than velocity components as was done in the previous subsection. The no-penetration condition $v(x,\pm 1, z) = 0$ implies both that $v_0 = 0$ and that the lowest $v$-mode goes as $\cos(\beta y)$. 
The five vertical modes used to represent the large-scale flow are illustrated in figure~\ref{fig:geometry}a.
The pressure field $p$ in \eqref{eq:ufull} that enforces incompressibility is represented in the form $p(x,y,z,t) = p_0(x,z,t) + p_1(x,z,t) \sin(\beta y)$. 

The retained modes in \eqref{eq:modes} represent the minimal set necessary to describe a three-dimensional vector field, and DNS of Wf confirms that these modes are the most energetic in turbulent bands. It is therefore reasonable to base our model on this minimal set. While $v_1$ is small in magnitude, it is essential to the flow dynamics. For the TKE field, $q_0$ represents the $y$-averaged turbulence and $q_1$ accounts for an asymmetry between the upper and lower halves of the domain. This is important for capturing the so-called {\em overhangs} at the edges of turbulent bands, where turbulence is not vertically uniform due to the mean shear \cite{coles1966progress,barkley2007mean,duguet2013oblique}. Substitution of \eqref{eq:modes} into \eqref{eq:full}, followed by a Galerkin projection, yields evolution equations for the fields. However, before proceeding, we must first consider the turbulence closures.

\subsection{Turbulence Closures} \label{subsec:closures}

The turbulence closures are based on DNS of Wf, described in \S\ref{subsec:gov_eq}, as well as four-mode simulations of Wf from previous work \cite{chantry2016turbulent}.
Each of the four fluctuation correlation terms \eqref{eq:restress} - \eqref{eq:transp} has an expansion in vertical modes; however, it is only the pseudo-dissipation $\varepsilon$ that will explicitly require two vertical modes.

\subsubsection{Reynolds Stress}

We start with the Reynolds stress tensor. The DNS reveals that only two components play a significant role in the dynamics, namely $-\langle u^\prime v^\prime \rangle$ and $-\langle u^\prime u^\prime \rangle$. In our truncated space of vertical modes, $-\langle u^\prime v^\prime \rangle$ has non-zero projection onto only the first vertical mode $\cos(\beta y)$. The truncation 
is supported by DNS results for a steady band in Wf at $Re = 140$, from which the root-mean-square (RMS) magnitude of the first vertical mode of $-\langle u^\prime v^\prime \rangle$ is, respectively, 10.2, 7.5, and 67.7 times larger than the second, third, and fourth vertical modes. 
The $-\langle u^\prime u^\prime \rangle$ component projects onto both the zeroth (constant in $y$) and first vertical modes, but from DNS, the zeroth mode is dominant (5.2 times larger than the first vertical mode for a steady band in Wf at $Re = 140$) and we consider only the zeroth mode. 
It is assumed that the amplitudes of these Reynolds stress components depend only on the TKE and have the form:
\begin{eqnarray}
    \mathcal{R}_{12} = \mathcal{R}_{21} = -\langle u^\prime v^\prime \rangle &=& A(q) \cos(\beta y), \label{eq:R12} \\
    \mathcal{R}_{11} = -\langle u^\prime u^\prime \rangle &=& -B(q).
\end{eqnarray}
This results in two Reynolds-stress forces -- one in the streamwise direction $x$, and one in the wall-normal direction $y$:
\begin{equation}
    \nabla \cdot {\bf \mathcal{R}} = \left(- A(q) \beta \sin(\beta y) -\partial_x B(q) \right) {\bf e}_x + \left(\cos(\beta y) \partial_x A(q) \right) {\bf e}_y.
\end{equation}
We have chosen the signs so that $A(q),B(q) > 0$. 

We now neglect the $-\langle u^\prime u^\prime \rangle$ stress, and take $B(q)=0$ for the remainder of this work. The reason we have included it in the discussion to this point is because this stress accounts for the largest effect that we have measured in DNS but that we do not include in the present model. This stress couples the $u_0$ field to $q_0$ and $q_1$. Its associated Reynolds-stress force acts to reduce the amplitude of $u_0$, while its associated production term increases $q_0$ (its effect is small compared to other production terms). We have briefly investigated the role of this stress in the model and have found that including it does not alter the model dynamics in a significant way. Therefore, for simplicity, we neglect it in the present model. Future studies may wish to include it. We find that taking $B(q)$ linear in $q$ is a reasonable approximation. 

\begin{figure}[ht]
	\centering
\includegraphics[width=0.9\textwidth]{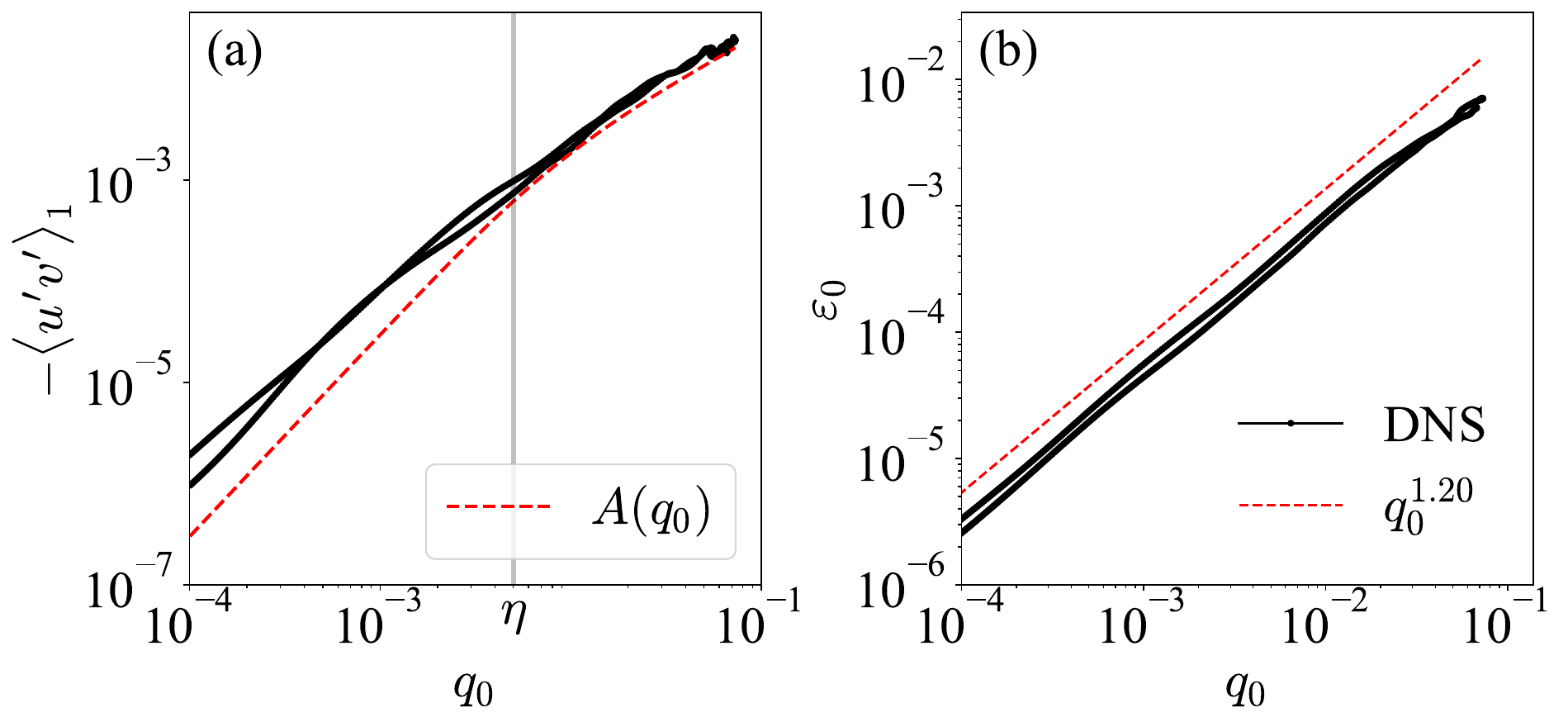}
\caption{
Functional dependence of (a) the first mode amplitude of $-\langle u^\prime v^\prime \rangle$ on $q_0$, and (b) the vertically averaged pseudo-dissipation rate $\varepsilon_0$ on $q_0$. Data is from a DNS of a steady band in Wf at $Re = 140$, with averaging along the band and over 1000 time units.
The black curve tracks the values of $-\langle u^\prime v^\prime \rangle$ or $\varepsilon_0$ with $q_0$ as the across-band direction is traversed. The maximum of $q_0$ occurs at the band centre, while small $q_0$ corresponds to the quasi-laminar regions between turbulent bands. Two lines appear because the averages extracted from DNS are not exactly symmetric about the band centre.  
See for example figure~\ref{fig:closures}b for the variation of $\varepsilon_0$ across the band direction.
In (a) the dashed line shows $A(q_0)$ for the model and the vertical grey line highlights $q_0 = \eta$, where $\eta=5\times 10^{-3}$.
In (b) the dashed line (displaced vertically for clarity) represent best fits of the DNS data.
} 
\label{fig:A_eps_vs_q}
\end{figure}

\begin{figure}[ht]
	\centering
\includegraphics[width=0.95\textwidth]{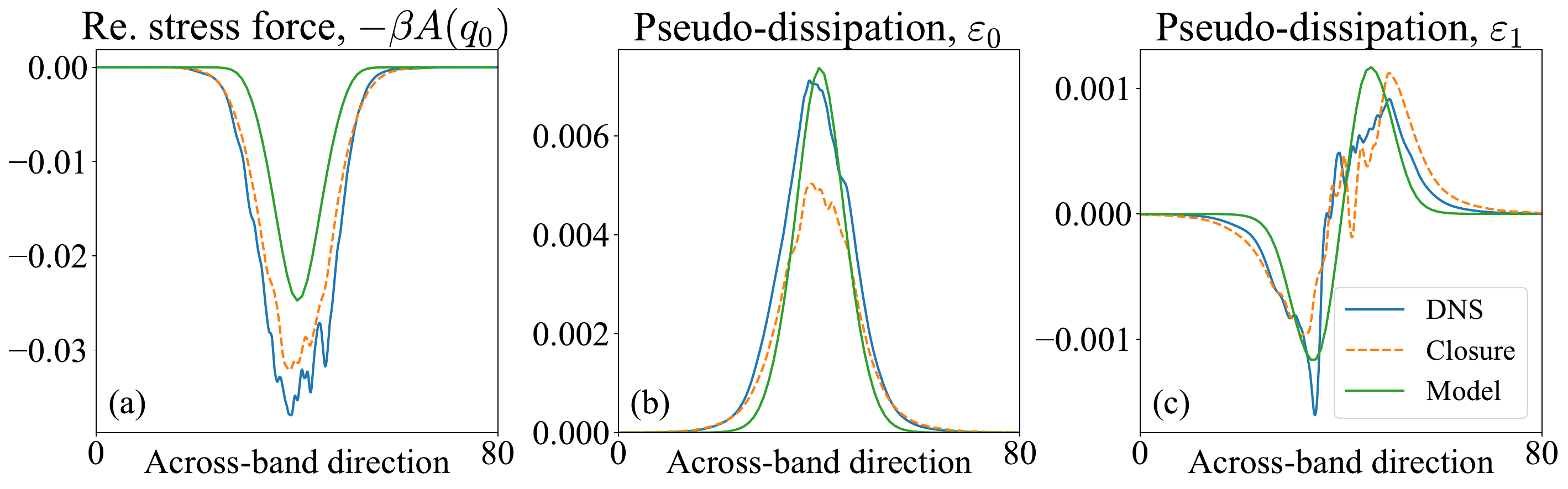} 
\caption{Comparison of Reynolds-stress force and pseudo-dissipation from DNS, model closures, and a model solution. Panel (a) shows the Reynolds-stress force in the $u_1$ mode.
DNS results are obtained from a steady band in Wf at $Re = 140$, with averaging along the band and over 1000 time units. 
The dashed curve shows the corresponding force using closure \eqref{eq:A_closure} with $q_0$ obtained from the DNS. The model curve (green) shows the force for a steady band in the model at $Re = 75$. (See \S \ref{subsec:params} for a discussion on the shift in $Re$.) Panels (b) and (c) show similar plots, but for the pseudo-dissipation rates $\varepsilon_0$ and $\varepsilon_1$, respectively.
} 
\label{fig:closures}
\end{figure}

Since $A(q)$ is the amplitude of an even mode in $y$, it depends on the even part of the TKE field, that is $q_0$ in our model expansion. The Reynolds-stress closure is thus specified by a function $A(q_0)$. We use the following functional form,
\begin{equation}
	A(q_0) = a ((q_0^2 + \eta^2)^{1/2} - \eta), \label{eq:A_closure}
\end{equation}
where $a$ and $\eta$ are parameters with $\eta$ small. Figure~\ref{fig:A_eps_vs_q}a compares $A(q_0)$ with data from DNS. For $q_0 \gg \eta, A(q_0) \approx a q_0$, meaning that $\mathcal{R}_{12}$ is simply proportional to the TKE.
For small $q_0$, that is, $q_0 \ll \eta$, we have $A(q_0) \approx a q_0^2 / 2 \eta$, and $A(q_0)$ goes to zero quadratically in $q_0$. This serves to cut off turbulent production at small $q_0$. The importance of this will be clear in the analysis of the model in \S\ref{sec:local_dynamics}.
While $A(q_0)$ underestimates the data for small $q_0$, and the Reynolds stresses do depend slightly on $Re$, we find that fitting the stresses to a more complicated functional form does not significantly alter the model dynamics. For simplicity, we choose not to account for these factors in the present model.
Figure~\ref{fig:closures}a compares DNS-measured values of the actual Reynolds-stress force across a turbulent band with closure \eqref{eq:A_closure}, as well as with the corresponding force from model simulations of a turbulent band (figure~\ref{fig:panels_small}a, $t = 25350$).


\subsubsection{Production}

The production $\mathcal{P}$ is just the contraction of the Reynolds stress tensor $\mathcal{R}$ and large-scale velocity gradient tenor $\nabla \bu$, followed by projection back onto the 
retained modes. Given our simplification of $\mathcal{R}$, the only terms appearing in the production are
$\mathcal{R}_{12} \partial_y u + \mathcal{R}_{21} \partial_x v$.
Using our expansions and projecting gives
$$
\mathcal{P} = 
\frac{1}{2}A(q_0) [\beta u_1 + \partial_x v_1].
$$
It is constant in $y$ and only couples to $q_0$.

It is worth emphasizing that the large-scale flow and TKE equations, \eqref{eq:ufull} and \eqref{eq:qfull}, are coupled only via the advective nonlinear term $\bu \cdot \nabla q$, 
which requires no modelling assumptions other than the vertical mode truncation, 
and via the Reynolds stress and production terms. As the model closure for the Reynolds stress also dictates the production, this closure completely determines the coupling between the large-scale flow $\bu$ and the TKE $q$. 

\subsubsection{Dissipation and Transport}
\label{sec:dissipation}
There are two closures left in the TKE equation, the turbulent pseudo-dissipation rate and the turbulent transport. We first focus on the pseudo-dissipation, which we explicitly expand as $\varepsilon = \varepsilon_0(x,z,t) + \varepsilon_1(x,z,t) \sin(\beta y)$.

As with $A(q)$, the even term in the pseudo-dissipation rate $\varepsilon_0$ depends only on $q_0$. DNS results show that the dependence of $\varepsilon_0$ on $q_0$ is approximately a power-law, with an exponent slightly greater than one (figure \ref{fig:A_eps_vs_q}b). In keeping with our goal of model simplicity, we model this dependence as linear. Tests confirm that a linear approximation is adequate (figure \ref{fig:closures}b). DNS results also support that the coefficient of proportionality depends on $Re$ approximately as $Re^{-1}$ throughout the transition region. Hence, our closure for $\varepsilon_0$ is
\begin{equation}
    \varepsilon_0(q_0;Re) = \frac{c}{Re} q_0, \label{eq:eps0}
\end{equation}
where $c$ is a model parameter. A similarly simple closure exists for $\varepsilon_1(q)$, that is $\varepsilon_1(q_1) = \kappa q_1$, where $\kappa$ does not depend on $Re$ (figure \ref{fig:closures}c). 

Turning to the transport term ${\bf T}$, DNS results and model tests indicate that only the zeroth vertical mode of the transport term plays a significant role in the turbulence budget, and thus in our model we ignore all but the zeroth vertical mode for this term. We invoke the gradient-diffusion hypothesis \cite[\S 10.3]{pope2000turbulent} which states that,
\begin{equation}
    {\bf T}(q_0;Re) = -\nu_T(q_0;Re) \nabla_H q_0.
    \label{eq:T_model}
\end{equation}
This is the simple physical assertion that (due to microscopic velocity and pressure fluctuations) there is a flux of $q_0$ down the gradient of $q_0$, with a transport coefficient, turbulent diffusivity $\nu_T$, itself depending on the $q_0$. 
The turbulent diffusivity is defined in a standard way to be proportional $q^2_0/\varepsilon_0$. (See Pope \cite[\S 10.3]{pope2000turbulent}.) Given the form for $\varepsilon_0 = c q_0 / Re$, equation \eqref{eq:eps0}, the turbulent diffusivity thus has the form
\begin{equation}
    \nu_T(q_0;Re) = d \,  Re \,  q_0,
    \label{eq:nu_T}
\end{equation}
where $d$ is a model parameter. 

\subsection{Quasi-static approximation for the $q_1$ dynamics}
\label{sec:q1}

We simplify the model through a series of assumptions on the $q_1$ field which allow us to express it as an explicit function of $q_0$ and the large-scale flow. Projecting the TKE dynamics \eqref{eq:qfull} into the $\sin(\beta y)$ mode gives the evolution equation for $q_1$ field. We assume $q_1$ adjusts on a fast timescale (quasi-static approximation) and that the advective derivatives in $q_1$ are small compared to those in $q_0$, as supported by DNS. The resulting evolution equation for the $q_1$ field is,
\begin{equation}
\cancel{\frac{\partial q_1}{\partial t}} + u_1\partial_x q_0 + w_1 \partial_z q_0
+ \cancel{u_0\partial_x q_1} + \cancel{w_0 \partial_z q_1} = -(2 \alpha  + \beta^2/Re) q_1 - \kappa q_1,
\label{eq:q1_time}
\end{equation}
where we have crossed out the terms that we ignore based on the assumptions above. 
As already noted, turbulent transport does not play a significant role in the $q_1$ dynamics and is not included. 

Our assumptions imply that the $q_1$ field adjust instantaneously, such that the linear dissipative terms in $q_1$ are always in balance with the source terms due to the advection of $q_0$.
Thus, in our model, the $q_1$ field is simply proportional to the dominant advective source driving it, and is given by 
\begin{equation}
q_1(u_1,w_1,q_0) = -\frac{(u_1\partial_x + w_1 \partial_z )q_0}{2 \alpha + \kappa + \beta^2/Re}.
\end{equation}

\subsection{Parameter fits}
\label{subsec:params}

\begin{figure}[ht]
	\centering
\includegraphics[width=0.9\textwidth]{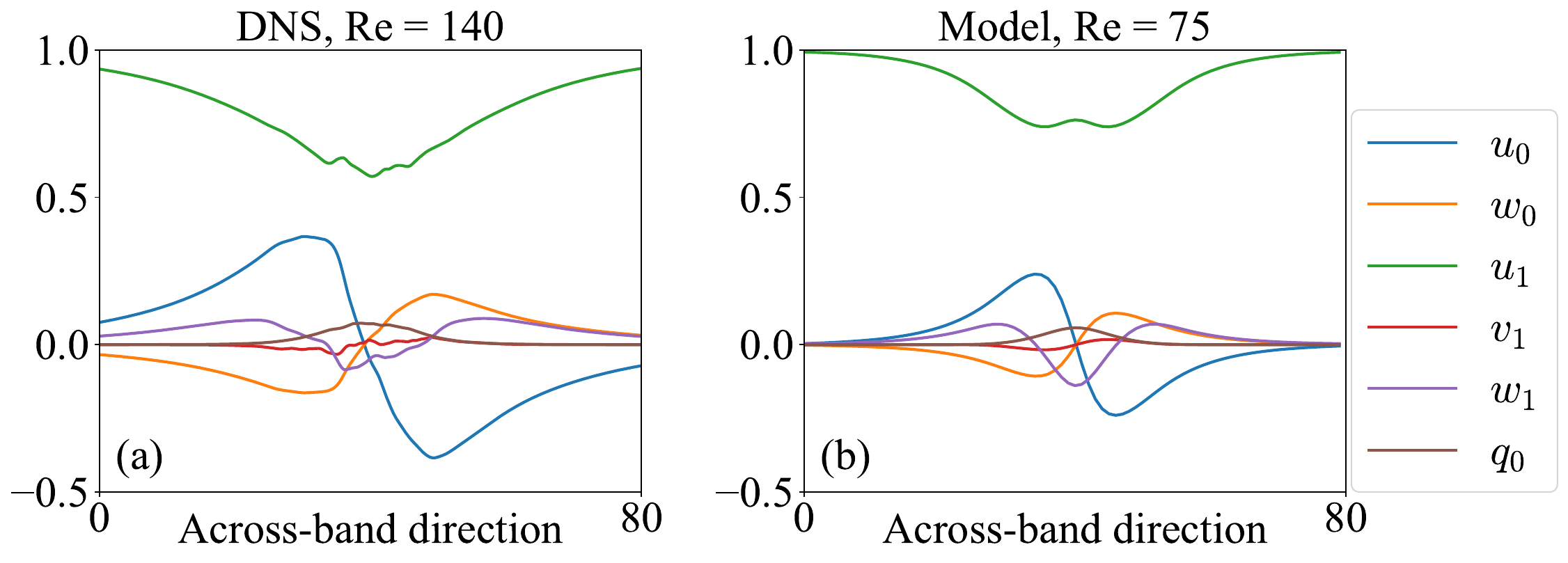} 
\caption{Across-band profiles of mode amplitudes from (a) DNS of a steady band in Wf at Re = 140, with averaging along the band and over 1000 time units, and (b) a model band at $Re = 75$. Although curves are plotted as a function of the across-band direction, the vectors have not been rotated, so that, for example, $u_0$ and $u_1$ still represent vertical modes of the streamwise velocity.} 
\label{fig:field_comp}
\end{figure}

Model parameters have been calibrated starting from a DNS of a turbulent band in Wf at $Re = 140$. 
From measured Reynolds-stress amplitudes, turbulent dissipation, and turbulent diffusion, we obtained a first estimate for model parameters. 
We subsequently refined values to ensure model simulations match reasonably well those of the DNS, as shown in figure \ref{fig:field_comp}, albeit with a necessary shift in $Re$ (discussed momentarily). We summarize here the most relevant aspects of the parameter fitting.

The Reynold stress and dissipation from the DNS of Wf are seen in figure~\ref{fig:A_eps_vs_q}.
The model parameters $a$ and $c$ associated with these effects,  \eqref{eq:A_closure} and \eqref{eq:eps0}, are particularly important as they dictate the onset of sustained turbulence in the model. (See \S \ref{sec:local_dynamics}.) We attempted to tweak $a$ and $c$ so that sustained turbulence first appears at Reynolds numbers close to those observed in Wf.
However, matching these Reynolds numbers results in the formation of sharp gradients in the model large-scale flow, which also leads to instability and unphysical phenomena.
We therefore settled on values of $a$ and $c$ such that model turbulent bands occur around $Re \simeq 75$. These $Re$ values are roughly half of the corresponding values in Wf \cite{chantry2016turbulent}. This shift to lower $Re$ is a common feature in modal truncations \cite{chantry2016turbulent,manneville2011modelling}. Indeed, turbulent bands occur in the range $50 \lesssim Re \lesssim 90$ in a four-vertical-mode truncation of Wf \cite{chantry2016turbulent,chantry_universal}, so it is sensible to aim for such Reynolds numbers in our model.  
Our study will focus on Reynolds numbers in the range $60$ to $100$, where the model exhibits the spatiotemporal dynamics of transitional turbulence. Outside the range the model produces only either fully laminar or fully turbulent solutions.

After calibrating $a$ and $c$, we performed simulations of turbulent bands using the model equivalent of the tilted domain technique \cite{barkley2005computational}, much like the DNS used for calibration.
Assuming bands are tilted at $24^\circ$ to the streamwise direction and are invariant in the along-band direction, the model reduces to a system of one-dimensional (1D) PDEs in the across-band coordinate, which we term the `1D tilted domain'. 
We performed a series of simulations and observed how the turbulent bands' width, amplitude, and disappearance to uniform flow depended on $Re$ and other model parameters. From these runs, we modified the rest of the parameters, in particular the turbulent diffusivity coefficient $d$, so as to best match the results found in the DNS. We did not attempt a more sophisticated parameter optimization beyond what is described here.

Throughout our experimentation in the tilted 1D domain, we observed steady bands and their transition scenario to be extremely robust to parameter changes. In the 2D model simulations, we also observed initial band formation to be a very robust phenomenon. However, experimenting with parameter values in 2D simulations also revealed various instabilities in the along-band direction and cases where the large-scale-flow amplitude is unrealistically large. 
Neglecting the Reynolds stress $\langle u^\prime u^\prime \rangle$, which plays a significant role in the budget of $u_0$, could be an issue, but tests including $B(q_0)$ did not solve the issue of band instabilities.
We believe that ultimately the issue is that the truncation of vertical modes 
results in too little dissipation in the large-scale flow, especially in the zero modes $u_0$ and $w_0$. Even after choosing parameter values that lower Reynolds numbers in the model, we found it necessary to increase the value of the drag coefficient $\alpha$ for the zero modes. This solves the issues of instabilities and the amplitude of the large-scale flow. We let $\alpha_0$ denote the drag coefficient for the zero modes and in practice use $\alpha_0 = 3 \alpha$. The introduction of $\alpha_0$ is our final modelling assumption.

\subsection{Summary}

In summary, the model is represented by six fields, $u_0, u_1, v_1, w_0, w_1, q_0$, corresponding to the  mode amplitudes of large-scale flow and TKE. The evolution equations of these fields in $(x,z,t)$ are obtained by substituting expansion \eqref{eq:modes} into the Reynolds average equations \eqref{eq:full}, using model closures for ${\bf \mathcal{R}}$, $\mathcal{P}$, $\varepsilon$, and $\textbf{T}$, using a quasi-static approximation for $q_1$, imposing incompressibility, and applying a Galerkin projection. 
The resulting evolution equations are
\begin{subequations}
\begin{align}
    \frac{\partial u_0}{\partial t} & + (u_0 \partial_x + w_0 \partial_z) u_0+ \frac{1}{2} (u_1 \partial_x + \beta v_1  + w_1 \partial_z) u_1 
    = -\partial_x p_0 + \left(\frac{1}{Re}\nabla^2_H - \alpha_0 \right) u_0,  \\
    \frac{\partial u_1}{\partial t} & + (u_0 \partial_x + w_0 \partial_z) u_1 + (u_1 \partial_x + w_1 \partial_z) u_0 
    = -\partial_x p_1 + \left(\frac{1}{Re}(\nabla^2_H - \beta^2) - \alpha \right) u_1 \nonumber \\
    & \hspace{6cm} + f - \uwave{\beta A(q_0)}, \\    
    \frac{\partial v_1}{\partial t} & + (u_0 \partial_x + w_0 \partial_z ) v_1 
    =  -\beta p_1 + \frac{1}{Re}(\nabla^2_H - \beta^2) v_1 + \uwave{\partial_x A(q_0)}, \\
    \frac{\partial w_0}{\partial t} & + (u_0 \partial_x + w_0 \partial_z) w_0 + \frac{1}{2} (u_1 \partial_x + \beta v_1 + w_1 \partial_z) w_1
    = -\partial_z p_0 + \left(\frac{1}{Re}\nabla^2_H - \alpha_0 \right) w_0, \\
    \frac{\partial w_1}{\partial t} & + (u_0 \partial_x + w_0 \partial_z) w_1 + (u_1 \partial_x + w_1 \partial_z) w_0 
    = -\partial_z p_1 + \left(\frac{1}{Re}(\nabla^2_H - \beta^2) - \alpha \right) w_1, \\    
    \frac{\partial q_0}{\partial t} & + (u_0 \partial_x + w_0 \partial_z) q_0 + \frac{1}{2} (u_1 \partial_x + \beta v_1 + w_1 \partial_z) q_1(u_1,w_1,q_0) \nonumber \\ 
    & = \frac{1}{2}\uwave{A(q_0)} [\beta u_1 + \partial_x v_1] -2\alpha q_0 - \uwave{\varepsilon_0(q_0;Re)} +
    \frac{1}{Re}\nabla^2_H q_0 + \uwave{\nabla_H \cdot \left(\nu_T(q_0;Re) \nabla_H q_0 \right) },
\end{align}
\label{eq:model_full}
\end{subequations}
where $f=\alpha + \beta^2/Re$, $\beta = \pi/2$, and 
$\nabla_H = (\partial_x, \partial_z)$.
The terms coming from the turbulence closure are highlighted by wavy underline. The only forcing of the large-scale flow by Reynolds stresses appears only via two terms. All other terms in the momenta equations arises just from modelling by Ekman friction and a Galerkin projection of the Navier-Stokes equations onto the retained modes.

The pressures $p_0$ and $p_1$ ensure incompressibility for each set of large-scale modes,
\begin{equation*}
    \partial_x u_0 + \partial_z w_0 = 0, \quad \partial_x u_1 - \beta v_1 + \partial_z w_1 = 0.
    \label{eq:div_full}
\end{equation*}
The equations are closed with the following,
\begin{equation}
    A(q_0) = a \, ((q_0^2 + \eta^2)^{1/2} - \eta), \quad
    \varepsilon_0(q_0;Re) = \frac{c}{Re} q_0, \quad
    \nu_T(q_0;Re) = d \,  Re \,  q_0. \label{eq:closures_turb} 
\end{equation}
\begin{equation}
    q_1(u_1,w_1,q_0) = -\frac{(u_1\partial_x + w_1 \partial_z) q_0}{2 \alpha + \kappa + \beta^2/Re}.
    \label{eq:closures_adv}
\end{equation}

Equations \eqref{eq:closures_turb} constitute the turbulence closures, expressing the dependence of Reynolds stress, pseudo-dissipation, and turbulent transport on turbulence amplitude $q_0$. Equation \eqref{eq:closures_adv} follows from the quasi-static approximation for the $q_1$ field. 

Model parameters have the values
\begin{equation}
    a = 0.3, \, \eta = 5\times 10^{-3}, \,  c = 9.65, \, d = 0.05, \, \kappa = 0.09, \, \alpha = 0.01, \, \alpha_0 = 0.03.
    \label{eq:parameters_full}
\end{equation}
The parameters $a$ and $\eta$ control the Reynolds stress closure, with $a$ setting its amplitude and $\eta$ setting the turbulence level below which the turbulence production rapidly decreases. 
Parameter $c$ sets the amplitude of turbulent dissipation associated with $q_0$;  $d$ is the coefficient for the diffusivity of $q_0$. Parameter $\kappa$ sets the turbulent dissipation for $q_1$, which for the model affects the amplitude of $q_1$ via the quasi-static approximation. Finally, $\alpha$ is the Ekman drag coefficient and $\alpha_0$ is the modified Ekman drag coefficient for the zeroth vertical modes.


\section{Model dynamics}

\subsection{Local dynamics}
\label{sec:local_dynamics}

We begin by considering the model's local dynamics -- 
the homogeneous dynamics in the absence of spatial derivatives. 
One may naturally think of these as the dynamics of horizontally uniform states with no $(x,z)$ dependence, such as fully laminar flow or fully turbulent flow, or one may think more loosely of these as the model dynamics within a small domain, a minimal flow unit \cite{hamilton1995regeneration, waleffe1997self}, in which large-scale spatial variation is not possible.  
From either perspective, the local dynamics are very informative of the core model behaviour.  

Dropping all spatial derivatives in the model equations, one immediately sees that the only fields that play a role in the local dynamics are $u_1$ and $q_0$. The other fields decouple and are either necessarily zero by incompressibility, as is the case for $v_1$, or they undergo exponential decay to zero, as in the case of $u_0, w_0$, and $w_1$. We let $u_1(x,z,t) \rightarrow u(t)$ and $q_0(x,z,t) \rightarrow q(t)$. Then the local dynamics are governed by the ordinary differential equations
\begin{subequations}
    \begin{align}
        \dot u & = \myD (1-u) - \beta A(q), \label{eq:ode_u} \\
        \dot q & = \frac{1}{2}\beta u A(q) - 2 \alpha q - \varepsilon_0(q; Re). \label{eq:ode_q}
    \end{align}
    \label{eq:odes}
\end{subequations}

These equations capture the essence of the interaction between the sinusoidal shear profile with amplitude $u$ and turbulent kinetic energy with amplitude $q$. They contain the two most important model closures: $A(q)$ modelling the Reynolds-stress amplitude and 
$\varepsilon_0(q; Re)$ modelling the dominant component of the turbulent pseudo-dissipation rate. (See equations \eqref{eq:R12}, \eqref{eq:A_closure},  and \eqref{eq:eps0}.)
The $1$ appearing in the expression $(1-u)$ in momentum equation \eqref{eq:ode_u} comes from the body force that acts against viscous decay and Ekman friction to drive $u$ to toward the laminar value $u=1$. The Reynolds-stress divergence, $-\beta A(q)$, is a force acting against the shear whenever $q > 0$. 
Recall that $A(0)=0$.
In the TKE equation \eqref{eq:ode_q}, TKE is generated by production, $\frac{1}{2}\beta u A(q)$, and dissipated by Ekman friction and the turbulent pseudo-dissipation rate.

\begin{figure*}
	\centering
	\includegraphics[width=0.9\textwidth]{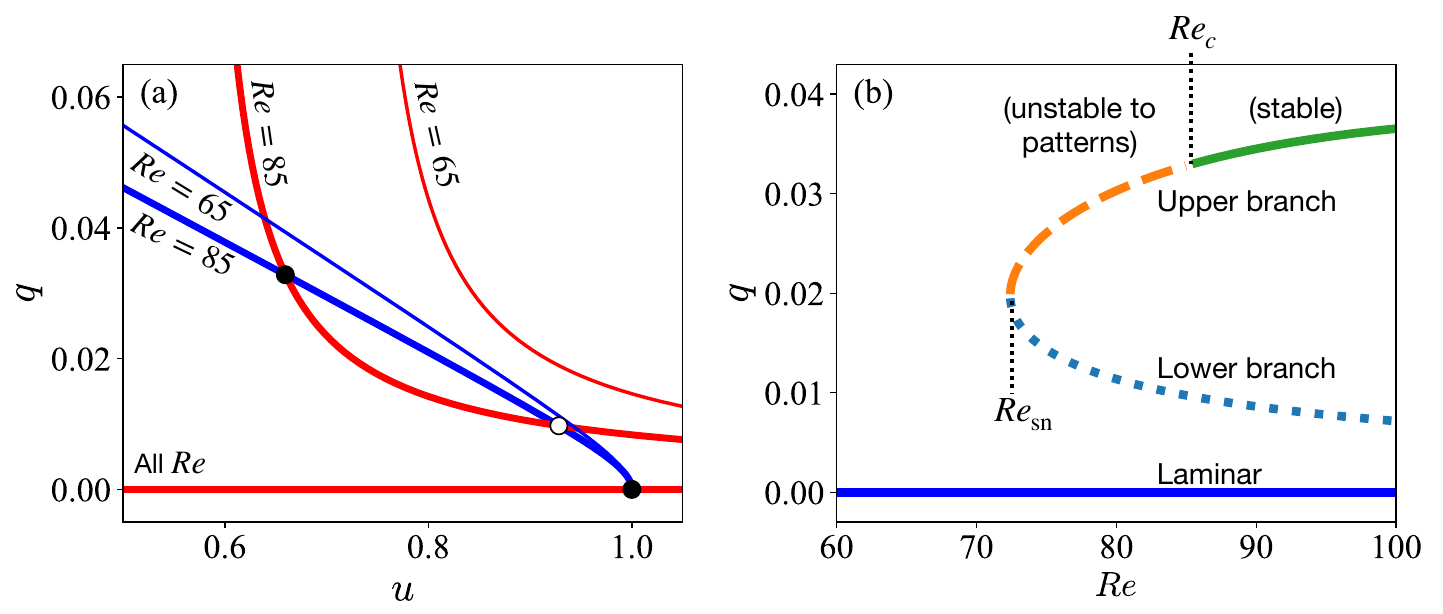} 
	\caption{
		(a) Nullclines and (b) bifurcation diagram for local dynamics. In (a), the $u$ nullclines, on which $\dot u = 0$, are shown in blue and the $q$ nullclines, on which $\dot q = 0$, are shown in red. Nullclines intersect at steady states of the local dynamics. At $Re=65$, the only steady state is laminar flow $(u_{ss},q_{ss})=(1,0)$, while at $Re=85$, there are two additional steady states. In (b), steady states are plotted as a function of $Re$. $Re_{\rm sn} = 72.4$ marks a saddle-node bifurcation in which upper- and lower-branch solutions emerge. Lower-branch solutions are unstable; upper branch solutions are unstable to patterns below $Re_c = 85.1$ and stable above. 
	} 
	\label{fig:local_dyn}
\end{figure*}

The local dynamics described by equations~\eqref{eq:odes} can be understood in terms of nullclines: curves in the $(u,q)$ phase plane on which $\dot u = 0$ and $\dot q = 0$. These are shown in figure~\ref{fig:local_dyn}a for two values of $Re$. Steady states of the local dynamics occur where the nullclines intersect and hence $\dot u = \dot q = 0$. We denote steady states by $(\uss,\qss)$. One steady state is laminar flow $(\uss, \qss) = (1,0)$ that exists for all $Re$. For $Re$ above $Re_{\rm sn} = 72.4$, the nullclines intersect at a pair of steady states with $\uss<1$ and $\qss > 0$, corresponding to  turbulent states. These states are shown in the bifurcation diagram in figure~\ref{fig:local_dyn}b, with turbulent states referred to as upper branch and lower branch depending on the value of $\qss$. These local dynamics are very similar to those in models exhibiting a Turing instability \cite{manneville2012turing, kashyap2025laminar}.

The linear stability of states will be addressed in full in \S~\ref{sec:stability}, but here we note that the laminar state is always linearly stable. Upper-branch states are linearly stable to spatially uniform perturbations, while the lower-branch states are unstable to such perturbations. Hence, the local dynamics given by equations~\eqref{eq:odes} exhibits bistability for $Re \ge Re_{\rm sn}$. While stable to uniform perturbations, the upper-branch state is unstable to spatial modulations, leading to patterns, below a critical value $Re_c = 85.1$, as indicated in figure~\ref{fig:local_dyn}b. 

Finally, the closures of the Reynolds stress $A(q)$ and the pseudo-dissipation $\varepsilon_0$ are central to the model. 
The dependence of $A(q)$ on $q$ gives a quadratic cut-off in production as $q \to 0$. Since $\varepsilon_0$ is linear in $q$, production will necessarily fall below dissipation at small $q$. This ensures the stability of the laminar flow with respect to small perturbations in $q$. 
Physically this rules out ``infinitesimal turbulence'', accounting for the fact that very small fluctuations (deviations from laminar flow), are not turbulent but rather are small laminar disturbances that we know do not sustain nonlinearly.


\subsection{Turbulent bands}
\label{sec:bands}

Going beyond the local dynamics, the model reproduces many features of turbulent bands observed in experiments and DNS of planar shear flows: plane Couette flow, Waleffe flow, and plane Poiseuille flow. We illustrate here some of the most significant dynamics captured by the model. 
For this we simulate the model equations on a doubly-periodic domain of size $L_x \times L_z$ using the pseudo-spectral code Dedalus \cite{burns2020dedalus}. Second-order Runge-Kutta time stepping is used with a time step $\Delta t = 0.04$, approximately 10 times larger than can be used in a corresponding DNS of the Navier-Stokes equations. 
The time step in both cases being set based on numerical stability.
A Fourier-spectral method with $3/2$ dealiasing is used with a resolution of one grid point per space unit, approximately ten times fewer than for a DNS. Based on the numerical time step and $(x,z)$ spatial discretizations alone, the model is $O(10^3)$ faster to simulate than the Navier-Stokes equations. 
Two types of initial conditions (ICs) are used to initialize simulations. A localized IC comprises a finite region of turbulence of length $25$ and width $14$ tilted at $24^\circ$ to the $x$ direction. 
The second IC is uniform turbulence $(u_0, u_1, v_1, w_0, w_1, q_0)(x,z,0) = (0,\uss,0,0,0,\qss)$. In all cases, the initial large-scale flow fields are additionally seeded with small-scale noise. 
(To account for turbulent fluctuations on the ``microscopic scale'', noise may be further included in simulations. We do not consider this here. The only source of randomness is the ICs.)%

\begin{figure*}
	\centering
	\includegraphics[width=\textwidth]{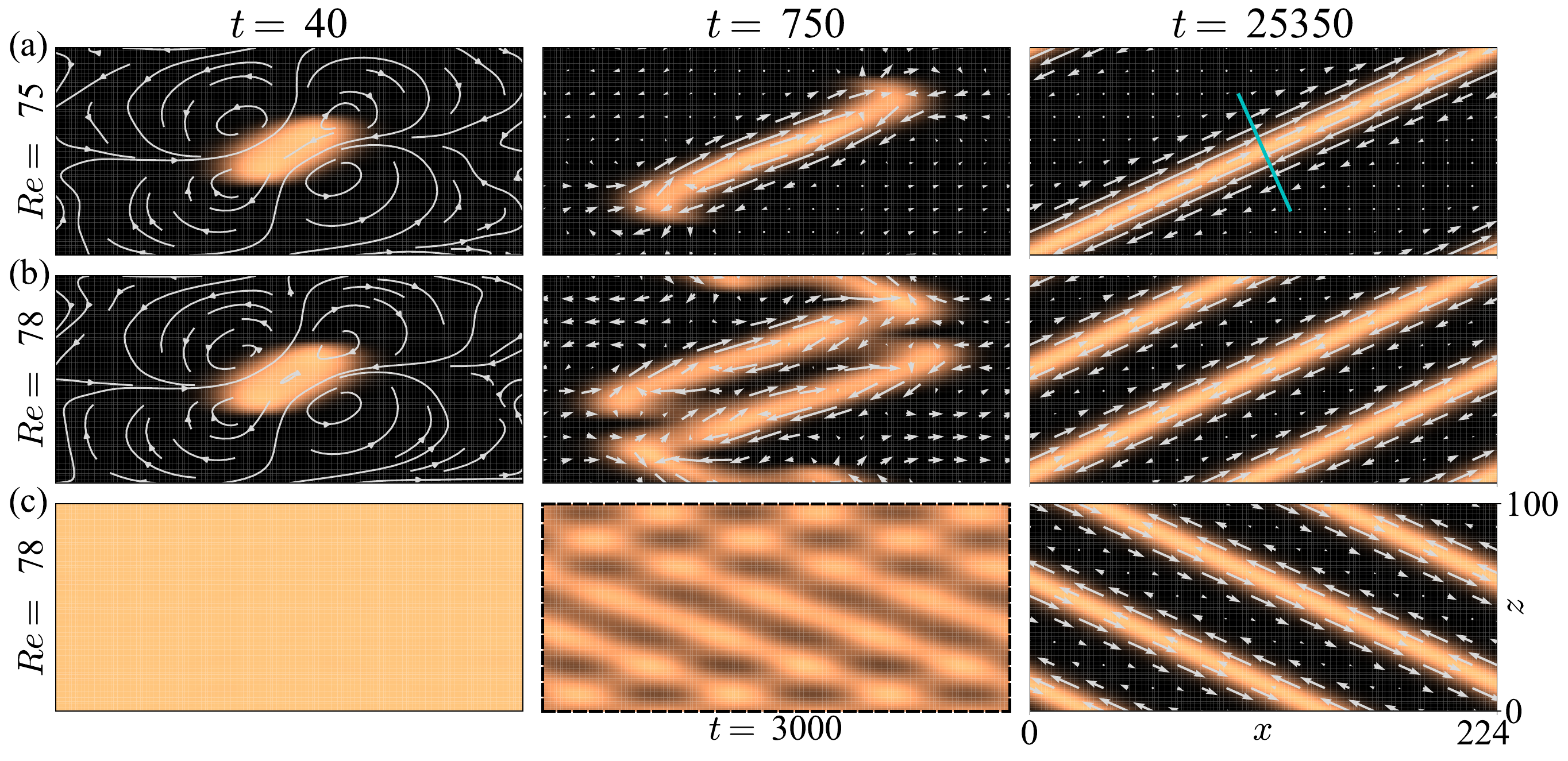}
	\includegraphics[width=\textwidth]{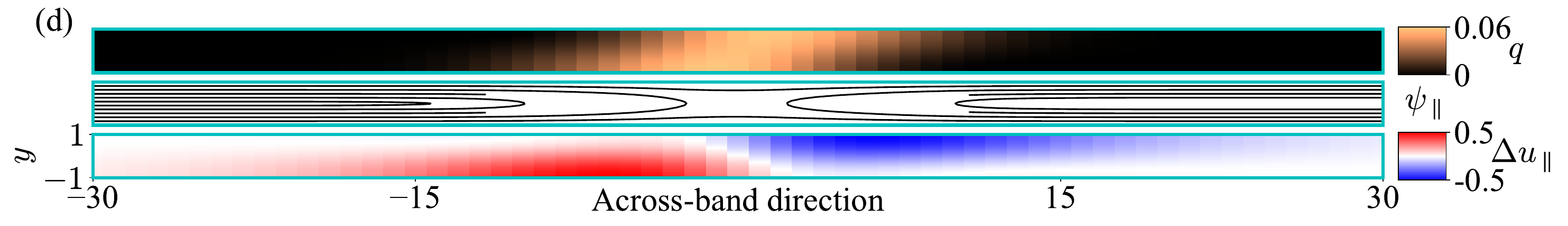}
	\caption{
		Representative model solutions starting from a localized initial condition for (a) $Re=75$ and (b) $Re=78$, and from a uniformly turbulent initial condition at $Re=78$ (c). Snapshot times are indicated, with the middle snapshot at $t=750$ for (a) and (b), but at $t=3000$ for (c).
        Visualized is the model field $q_0$, representing the vertically-averaged turbulent kinetic energy. Vectors and streamlines show the vertically-averaged large scale flow. 
        Panel (d) shows a vertical slice of the reconstructed turbulence field $q$ (top panel) and large scale flow velocities (bottom two panels), taken across the final steady band in (a), as shown with a cyan line. The middle panel in (d) shows the two-dimensional in-plane flow, or equivalently contours of the streamfunction $\psi_\parallel$ for this flow, while the bottom panel shows the along-band flow visualized as a deviation from the laminar flow, $\Delta u_\parallel \equiv u_\parallel - u_{lam \parallel}$, where $\parallel$ is the direction parallel to the band.
	} 
	\label{fig:panels_small}
\end{figure*}

\begin{figure*}
	\centering
	\includegraphics[width=\textwidth]{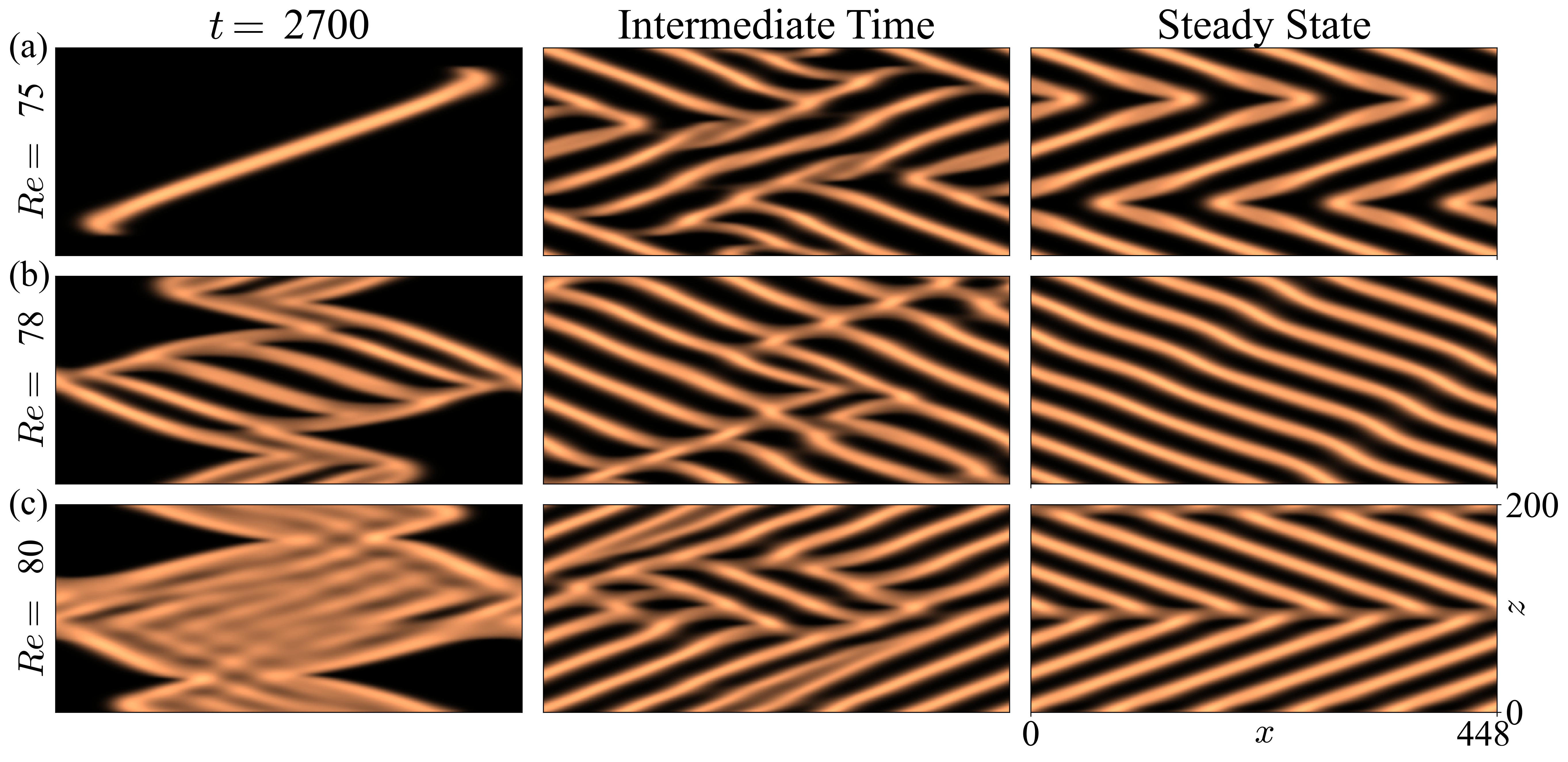} 
	\caption{
		Illustrative model dynamics in large domains. Temporal evolution of turbulent bands for three values of $Re$, all initialized with the same localized initial condition. (a) $Re = 75$, (b) $Re = 78$, and (c) $Re = 80$ \cite[Movies S9-S11, respectively]{SM}. The colour represents the amplitude of $q_0$, the vertically-averaged turbulent kinetic energy.
	} 
	\label{fig:panels_big}
\end{figure*}

Figure \ref{fig:panels_small}a shows a simulation at $Re = 75$ starting from the localized IC. 
During the initial evolution, 
a large-scale, quadrupolar velocity field is established. 
Such fields are well documented \cite{lundbladh1991direct,schumacher2001largescaleflow,lagha2007largescaleflow,duguet2013oblique,chantry2016turbulent,wang2020quadrupolar,kashyap2020far,couliou2016spreading,klotz2021experimental}, and reproducing them is an important validation of the model. 
The turbulent patch elongates through the growth of its two tips.  They eventually join and the system settles into a single steady, straight turbulent band 
tilted at $\theta = 24^\circ$ to the streamwise direction \cite[Movies S1-S2]{SM}. While the final angle of the band to the streamwise direction is constrained by the domain, during the initial band elongation (figure~\ref{fig:panels_small}a, $t=750$), the angle is not far from $\theta = 24^\circ$ (see also figure~\ref{fig:panels_big}a, $t=2700$, discussed below). Furthermore, doubling the domain size in the spanwise direction still results in $24^\circ$ bands \cite[Movie S8]{SM}.
This is consistent with numerous observations of bands where angles typically line in the range $24^\circ \lesssim \theta \lesssim 45^\circ$ \cite{tuckerman2020patterns,chantry_universal,prigent2003long,duguet2010formation,duguet2013oblique,ShimizuPRF2019,kashyap2022linear,klotz2022phase,xiong2015turbulent,xiao2020growth}.
The large-scale flow directed along the edges of the turbulent bands is typical of that observed in experiments and DNS \cite{coles1966progress,barkley2007mean,duguet2013oblique,gome2023patterns1,marensi2022dynamics}. It is worth emphasizing that the large-scale velocity vectors seen in the midplane are given by $u_0(x,z) {\bf e}_x + w_0(x,z) {\bf e}_z$.


Figure \ref{fig:panels_small}b shows a simulation at $Re = 78$. At this slightly higher value of $Re$, we observe more growth at the tips and also lateral splitting of the band, eventually resulting in two identical steady bands within the domain \cite[Movies S3-S4]{SM}. Figure~\ref{fig:panels_small}c shows the $Re = 78$ case again, but now starting from the uniform IC. Initially, there is competition between symmetrically related orientations (the system is symmetric under spanwise reflections, $z \to -z$). Eventually, one orientation emerges and a single pair of steady bands forms \cite[Movie S5]{SM}. 
Both lateral splitting \cite{barkley2005computational, duguet2010formation,manneville2012growthspot,marensi2022dynamics,lu2022growth} and band competition \cite{prigent2002large,prigent2003long,duguet2010formation,chantry_universal,klotz2022phase,kashyap2020flow,kashyap2022linear} are well-documented phenomena. 

Figure~\ref{fig:panels_small}d shows a vertical slice of the reconstructed turbulence field $q$ (top panel) and large-scale velocities (bottom two panels), taken across the final steady band at $Re = 75$ (cyan line in figure~\ref{fig:panels_small}a, t = 25350). The plots are strikingly similar to vertical slices in DNS of turbulent bands \cite{barkley2007mean,chantry2016turbulent,gome2023patterns1,tuckerman2020patterns}. The top panel highlights the so-called overhang turbulent regions \cite{coles1966progress,barkley2007mean,duguet2013oblique} that are generated through advection of the turbulent field by the mean shear and captured by the $q_1$ field in the model. The middle panel visualizes the streamlines of the large-scale flow in the slice plane and the bottom panel shows contours of the well-documented along-band flow (through the slice plane)
\cite{barkley2007mean, chantry2016turbulent, tuckerman2014turbulent}. The structure in the wall-normal direction seen in figure~\ref{fig:panels_small}d highlights that, while model fields are functions of $(x,z)$, they describe non-trivial, three-dimensional flows very much like those seen in DNS. 

We now survey the spatiotemporal dynamics of bands in the 1D tilted domain, such that bands are at a fixed angle and independent of the along-band direction. 
We do so by working in coordinates $(\xrot, \zrot)$ rotated by $\theta=24^\circ$ with respect to the $(x,z)$ coordinates and setting to zero all derivative in the along-band direction, $\xrot = x \cos\theta + z \sin\theta$.
Restricting to such solutions, the model reduces to a system of 1D PDEs in the across-band coordinate (Appendix~\ref{app:tilted}).
While restricting the band angle eliminates fully two-dimensional phenomena, such as seen in the early times in figure~\ref{fig:panels_small}, it provides insights into some of the key spatiotemporal features of turbulent bands.

\begin{figure}
	\centering
	\includegraphics[width=\textwidth]{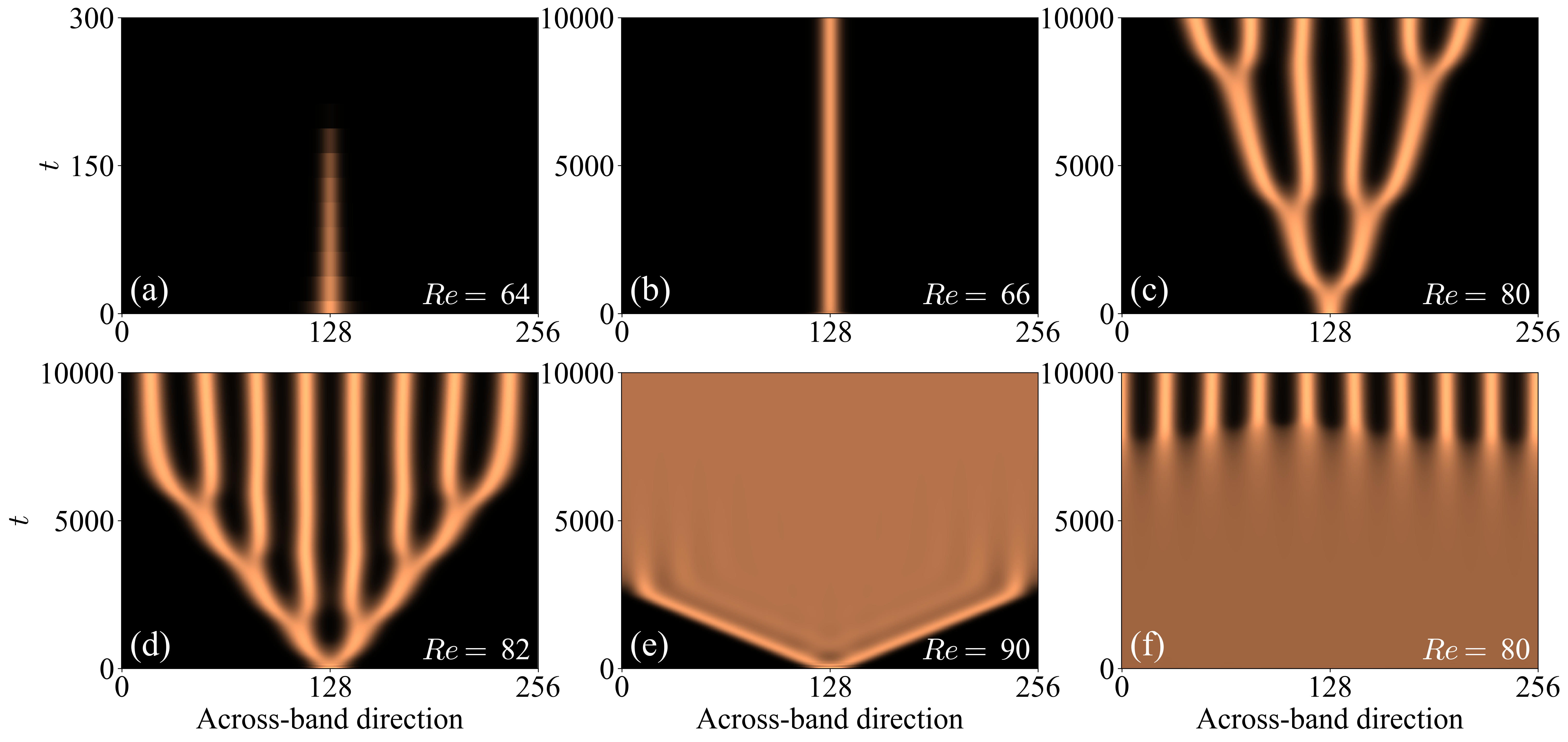} 
	\caption{Spatiotemporal dynamics of 1D bands at a fixed angle $\theta=24^\circ$, for various $Re$ indicated in the panels. Visualized is the turbulent field $q_0(z',t)$, where $z'$ is the across-band coordinate.
    (a)-(e) Dynamics starting from a localized turbulent patch illustrating the following behaviour: (a) relaminarization, (b) single, localized turbulent band,  (c) and (d) expanding turbulence in the form of band splitting or patterned slug, (e) expanding turbulent slug. (f) Illustration of band formation from uniform turbulence. Panels (c) and (f) correspond to the same conditions, just different initial conditions. Beyond the time shown in (c) the system develops eight turbulent bands.
	} 
	\label{fig:panels_1D}
\end{figure}

Figure~\ref{fig:panels_1D} shows spatiotemporal diagrams at representative values of $Re$, with the band angle fixed at $\theta=24^\circ$. Simulations are initiated from a localized IC (a steady band solution at $Re = 66$), except in figure~\ref{fig:panels_1D}f where the simulation is initiated with a uniform turbulent IC. 
At sufficiently low $Re$, turbulent bands are not sustainable and localized initial states rapidly decay, as illustrated at $Re=64$. Note the timescale of panel figure~\ref{fig:panels_1D}a in comparison with the other panels.
Starting at $Re \simeq 66$, steady bands are sustained in the form of localized, isolated structures. This is well below the onset of bistability in the homogeneous dynamics in figure~\ref{fig:local_dyn}.

At larger $Re$, figures~\ref{fig:panels_1D}c-\ref{fig:panels_1D}e, an initially localized turbulent patch spreads laterally. At $Re=80$ and $Re=82$, the spreading occurs via what is known as splitting, and the resulting asymptotic states are regular periodic patterns of turbulent bands. (Beyond time shown in figure~\ref{fig:panels_1D}c, the system reaches a steady pattern with eight turbulent bands.) At $Re=90$, turbulence expands rapidly in what is referred to as a slug \cite{wygnanski1973transition, darbyshire1995transition, barkley2015rise}. The asymptotic state is uniform turbulence. As can be seen, the expansion rate is an increasing function of $Re$. The final panel, figure~\ref{fig:panels_1D}e, shows the formation of bands from uniform turbulence at $Re=80$, the same $Re$ as in figure~\ref{fig:panels_1D}c. This demonstrates explicitly the instability of the uniform turbulent state at this $Re$. 


Finally, returning to the 2D model dynamics, we show in figure~\ref{fig:panels_big} some of the model’s behaviour in larger domains---twice the size in each horizontal dimension compared to figure~\ref{fig:panels_small}---starting from the localized ICs. While the initial evolution (first column of figure~\ref{fig:panels_big}) resembles that of runs in figure~\ref{fig:panels_small}, the larger domain better captures the complex process of turbulence proliferation. The single band in figure~\ref{fig:panels_big}a elongates through the growth of its two tips, as before, but no longer simply reconnects with itself upon reaching the boundaries. Instead, it produces a complex web of chaotic bands \cite[Movie S9]{SM}. 
At slightly higher $Re$, proliferation changes markedly due to band splitting (figure~\ref{fig:panels_big}b). Turbulence expands in all directions---albeit anisotropically---via the continuous generation of new bands through splitting \cite[Movie S10]{SM}. This spreading behaviour is strikingly similar to that observed in plane Couette flow \cite{duguet2010formation,marensi2022dynamics}.
At yet higher Reynolds number (figure~\ref{fig:panels_big}c), the initial spreading occurs more like a front of nearly uniform turbulence, which subsequently develops laminar gaps, resulting in bands \cite[Movie S11]{SM}. For all cases, intermediate times reveal long-lived complex dynamics as the system adjusts toward equilibrium. During this phase, bands of different orientations compete (middle column of figure~\ref{fig:panels_big}).
Not all combinations of ICs, Reynolds number, and domain size lead to simple bands. We observe a variety of other steady states, such as criss-crossing bands \cite[Movie S6]{SM}, non-uniform bands (figure~\ref{fig:panels_big}b), and instances featuring simple bands of both orientations separated by domain boundaries (figure~\ref{fig:panels_big}c) \cite{prigent2002large,prigent2003long}. 
Some runs never reach a steady state: bands nearly form, become unstable, break up, and re-form in a repeating cycle. The dynamics of the model are rich, and there is much more to explore, but we leave this for future work.


\section{Linear stability \label{sec:stability}}

The model provides a powerful means to analyze band formation in ways that experiment and DNS do not have access to. In particular, with the model we can directly address the transition from uniform turbulence to turbulent bands via linear stability analysis. We develop this approach over the next two sections. In this section, we presents numerical results and establish the fidelity of long-wavelength approximation to the linear stability equations. In the next section, we derive a bound on the critical angle for the onset of patterns in the long-wavelength approximation. In this section, we refer to the approximate system as the `long wavelength' system, and the model without any approximations as the `full' system.

\subsection{Stability analysis of the full system}

Recall that the uniform turbulent state in the model is a steady state of the form 
$(u_0, u_1, v_1, w_0, w_1, q_0) = (0,\uss,0,0,0,\qss)$, where $\uss$ and $\qss$ satisfy equations~\eqref{eq:odes} after dropping the $t$ derivatives. 
Straightforward linearization of the model equations about a uniform turbulent state gives an eigenvalue problem for linear modes of the form $\hat{u}_0 \exp(ik_x x + ik_z z + \sigma t)$ 
and similarly for $\hat{u}_1, \hat{v}_1, \hat{w}_0, \hat{w}_1, \hat{q}_0$, where $\sigma$ is the temporal growth rate and $k_x$ and $k_z$ are streamwise and spanwise wavenumbers of the perturbation. 
The linearized model equations are stated in full in appendix \ref{app:lin_stab} and details of their derivation are found in the electronic supplementary material \cite{SM}. 

For a given value of $Re$ and pair of wavenumbers $(k_x,k_z)$, the linearized equations become a simple matrix eigenvalue problem that can be written
\begin{equation}
   \sigma {\bf \hat x} = {\bf M} {\bf \hat x} 
\label{eq:Matrix}
\end{equation}
where ${\bf \hat x} = (\hat{u}_0, \hat{u}_1, \hat{v}_1, \hat{w}_0, \hat{w}_1, \hat{q}_0)^T$ and $\bf M$ is a $6 \times 6$ matrix. This equation is easily solved numerically for the eigenvalues $\sigma$, which can then be ordered by real part. Positive real part correspond to linear instability.
Stability at zero wavenumber, $(k_x,k_z) = (0,0)$, is equivalent to stability of the local dynamics, given by the linearization of ~\eqref{eq:odes}. 

Stability of the uniform states is indicated in the bifurcation diagram of figure~\ref{fig:local_dyn}b. The laminar state is linearly stable for all $Re$, independently of $(k_x,k_z)$. The lower-branch turbulent states are unstable. 
The focus is on possible instability of the upper-branch turbulent states.
At large $Re$ the upper branch is linearly stable, as expected since uniform turbulence is the observed state in model simulations. 
With decreasing $Re$, a positive real eigenvalue first appears at a critical Reynolds number $Re_c = 85.1$. 
Hence, for $Re < Re_c$ uniform turbulence is unstable to spatial modulations, as  indicated by the dashed portion of the upper branch in figure~\ref{fig:local_dyn}b, and as seen in figures~\ref{fig:panels_small}c and \ref{fig:panels_1D}f.


\begin{figure}
	\centering
	\includegraphics[width=0.9\textwidth]{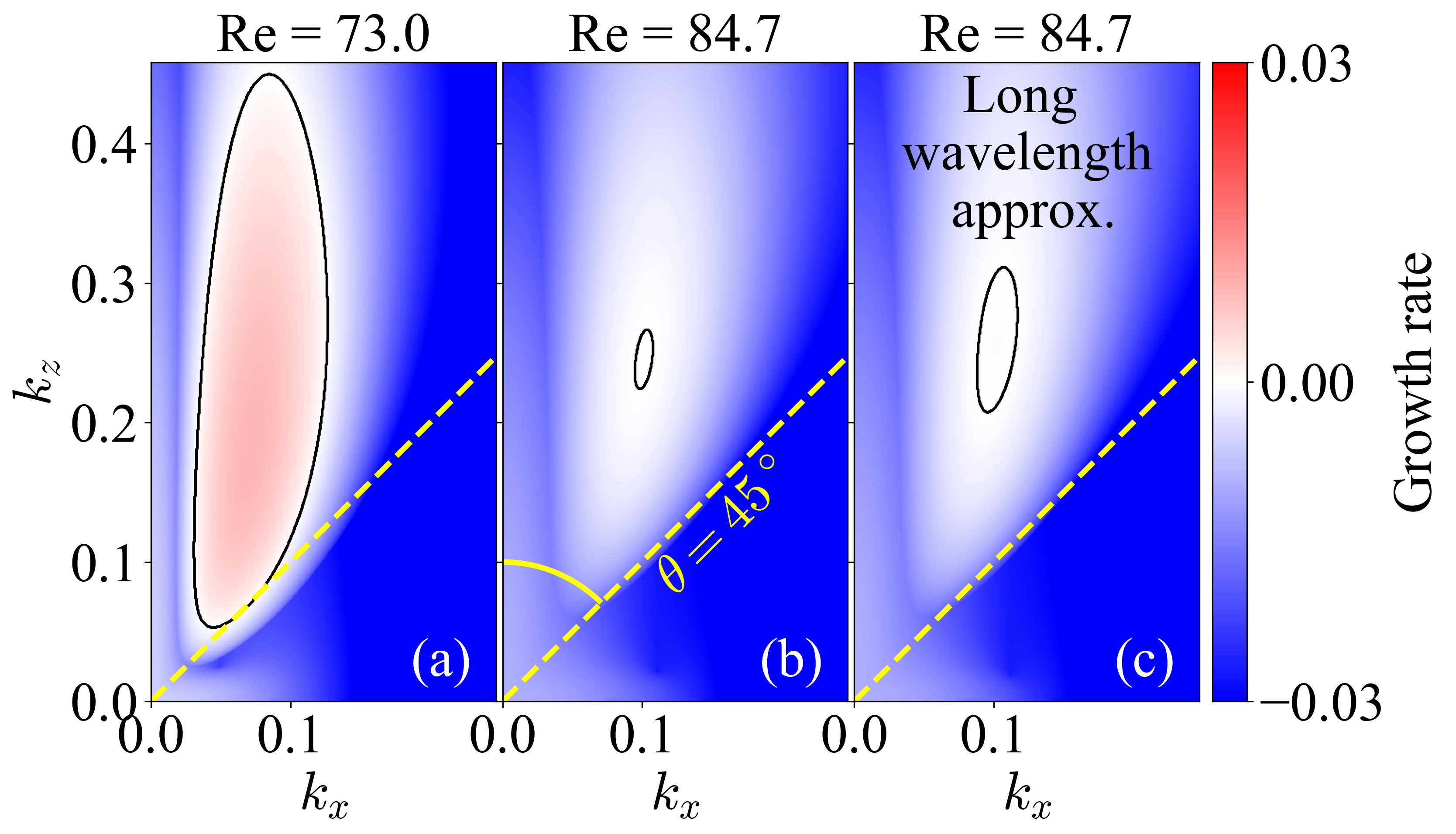}
	\caption{Maximum growth rate for linear perturbations to the upper-branch turbulent state at a Reynolds number (a) just above the onset of bistability, and (b) just below $Re_c$ where a positive growth rate appears. The solid black line represents the neutral stability curve. (c) Maximum growth rates of the simplified linearized equations under the long-wavelength assumption $(k\ll \beta)$.
	} 
	\label{fig:stability}
\end{figure}

Figure~\ref{fig:stability} shows the maximum growth rates in the $(k_x, k_z)$ plane for two values of $Re$: one just above the saddle-node marking the 
appearance of the upper branch and one just below $Re_c$ where instability sets in. 
A mode with wavenumbers $(k_x, k_z)$ corresponds to a stripped pattern with wavelength $\lambda = 2\pi/k = 2\pi/\sqrt{k_x^2+k_z^2}$ and  
angle $\theta$ to the streamwise direction, where $\tan \theta = k_x/k_z$.
At the onset of instability, $Re = Re_c$, the critical values of the wavelength and angle are: $\lambda_c = 23.6$ and $\theta_c = 22.5^\circ$. With decreasing $Re$, the unstable region enlarges and the fastest growing mode shifts to larger wavelength and slightly lower angle ($\lambda \approx 32$ and $\theta \approx 21.5^\circ$ at $Re = 73$). These results align very well with evidence of a linear instability of uniform turbulence in plane Poiseuille flow at a critical angle of $\theta_c \approx 23^ \circ$ \cite{kashyap2022linear}. 
Nonlinear model simulations of the 1D tilted domain setup show that the bifurcation to bands at $Re_c$ and $\theta_c$ is supercritical. That is, modulation strength grows continuously from zero as $Re$ is decreased below $Re_c$ with $\theta$ fixed at $\theta_c$. The transition is not hysteretic. 
The bifurcation in the model is not always supercritical, however; for some fixed angles and wavelengths, the transition from uniform turbulence to bands as $Re$ is decreased is subcritical. This is will discussed in a separate publication. 

\subsection{Long-wavelength simplification}
 
Turbulent bands have wavelengths that are an order of magnitude larger than the wall spacing. Equivalently, the ratio of horizontal wavenumbers $(k_x, k_z)$ to the wall-normal wavenumber $\beta$ is small. We can exploit this to remove small, unimportant terms from linear stability equations while maintaining all the essential balances. This will simplify the algebra and thereby clarify the analysis to follow in \S \ref{sec:crit_angle}.

Formally, we assume $k_x/\beta \ll 1$ and $k_z/\beta \ll 1$, or for simplicity, $k/\beta \ll 1$, so that terms order $(k/\beta)^2$ terms are negligible compared with order $(k/\beta)$, and larger, terms.
For example, the three-dimensional spectral Laplacian $-(k^2 + \beta^2)$ is approximated as  $-\beta^2$.
We relegate the details to appendix \ref{app:lin_stab}, and to \S \ref{sec:crit_angle} where we take up the stability equations explicitly. At present, it is sufficient to understand that the long-wavelength approximation slightly alters the stability matrix $\bf M$ in equation~\eqref{eq:Matrix}.
We do not introduced separate notation for the stability matrix, nor eigenvalues and vectors, for the approximate system.

Making the long-wavelength approximation, the critical values for the onset of patterns change slightly
to: $Re_c = 93.3$, $\lambda_c = 20.1$ and $\theta_c=21.5^\circ$.
The dominant effect of the approximation is a shift in the critical Reynolds number to a slightly larger value. 
Figure \ref{fig:stability}c shows the leading eigenvalue of the approximate system for the same conditions as in figure \ref{fig:stability}b. The region of instability in $(k_x,k_z)$ is slightly larger under the approximation, which can be accounted for by the shift in $Re_c$, but otherwise plots \ref{fig:stability}b and \ref{fig:stability}c are very similar. 
We now proceed to analyse the transition to patterns in the simplified stability equations. 

\section{Analysis of the critical point}

We show explicitly that, in this long-wavelength approximation, the selected angle of turbulent bands at onset, the critical angle $\theta_c$, is bounded by  $0 < |\theta_c| < 45^\circ$. 

\subsection{Infinitesimal balances}
\label{sec:balances}

The analysis can continue by focusing on the linear system at the onset of the instability, in other words, the case of marginal stability where $\sigma = 0$. This is both the most interesting case, as it corresponds to a bifurcation point, and also the most analytically tractable case, because with $\sigma = 0$ the stability equations become algebraic equations. 
In the long-wavelength approximation, these equations are
\begin{subequations}
\begin{align}
{\rm streamwise~(bulk):} & &
\underA{
\frac{1}{2} i k_x \uss \hat{u}_1 + \frac{1}{2} \beta \uss \hat{v}_1} 
& = 
\underP{\beta \uss \frac{k_x^2}{k^2} \hat{v}_1}
\underD{- \alpha_0 \hat{u}_0},
\label{eq:hatu0_4} \\
{\rm streamwise~(shear):} & &
\underA{i k_x \uss \hat{u}_0} 
& =  
\underD{
-(\alpha + \beta^2/{Re}) \hat{u}_1} 
\underR{- \beta \dA \hat{q}_0 },
\label{eq:hatu1_4}  \\
{\rm wall ~ normal:} & &
0 & = 
\underD{- (\alpha + \beta^2/{Re}) \hat{v}_1}
\underR{- ik_x \dA \hat{q}_0},
\label{eq:hatv1_4} \\
{\rm TKE:} & &
0 & = \frac{1}{2} \beta \Ass \hat{u}_1 + \Lhat \hat{q}_0.
\label{eq:hatq0_4}
\end{align}
\label{eq:hat_4}
\end{subequations}
This is a closed system for the momenta and energy balances that must hold for an infinitesimal perturbation, $(\hat u_0, \hat u_1, \hat v_1, \hat q_0)$, to be neutral.

Thus far we have relegated details of the model linearization to appendix \ref{app:lin_stab} and the electronic supplementary material. Now that we will explicitly present the equations, we take the time to discuss the physical meaning of the equations and the various terms that appear.
Equations \eqref{eq:hatu0_4} and \eqref{eq:hatu1_4} correspond to the equations for $\hat{u}_0$ and $\hat{u}_1$, respectively, and are streamwise momenta balances in the modes $\sin(0y)$ and $\sin(\beta y)$, 
naturally thought as the streamwise "bulk" and "shear" modes.
Equations \eqref{eq:hatv1_4} and \eqref{eq:hatq0_4} are wall-normal momentum and TKE balances, respectively. 
The spanwise-flow perturbations, $\hat{w}_0$ and $\hat{w}_1$, do not appear in equations \eqref{eq:hat_4}, and hence we may analyse bifurcations without explicit reference to the spanwise momentum balances. 

Terms on the left-hand-side result from linearization of the advective nonlinearity about the homogeneous steady state $(u_0, u_1, v_1, w_0, w_1, q_0) = (0,\uss,0,0,0,\qss)$. Factors of $1/2$ in \eqref{eq:hatu0_4} result from Galerkin projection onto the mean.
The pressure term enforces incompressibility of the infinitesimal flow perturbation and is calculated in the electronic supplementary material \cite{SM}.
Although the spanwise flow components do not appear in equations \eqref{eq:hat_4}, their presence is necessary for determining the pressure term. 
The dissipation terms, viscous and Ekman friction, are straightforward as these are linear effects. The Reynold stresses 
are modelled with $A(q_0)$, and when linearized about $\qss$ become $\dA \hat q_0$, where $\dA \equiv dA/dq(\qss)$. The Reynold-stress forces in \eqref{eq:hatu1_4} and \eqref{eq:hatv1_4} include a pressure projection to divergence-free form.

The terms in the TKE balance \eqref{eq:hatq0_4} are more involved, but as we shall see, the details are mostly unimportant for the analysis to follow. The large-scale flow field enters the TKE budget only through the production $\mathcal P$ (the product of Reynolds stress and velocity gradient). When linearized about $\qss$, this gives a single term involving the infinitesimal flow perturbation $\hat u_1$, where $\Ass = A(\qss)$ is the Reynolds stress of the homogeneous state. All other terms in the linearization give the factors of the TKE perturbation $\hat q_0$. Explicitly the terms are
\begin{equation}
    \Lhat = \underbrace{
    \frac{1}{2}\beta \uss \dA 
    - 2 \alpha 
    - \deps}_{\Lzero(Re)}
    - k^2 \underbrace{\left (
    \frac{1}{Re} 
    + \nu_T(\qss;Re) 
    + \frac{\uss^2 \sin^2 \theta}{2(2 \alpha + \kappa + \beta^2/Re)}
    \right )}_{\Lplus(\theta,Re)}.
\label{eq:Lop_ktheta}
\end{equation}
We have separated the $k$-independent terms from those proportional to $-k^2$, and used that  $k_x^2 = k^2 \sin^2 \theta$. The three terms in $\Lzero$ are the linearization of the production, the Ekman friction, and the dissipation, respectively, where $\varepsilon' = d \varepsilon_0/d q_0 (\qss)$. These are necessarily the terms that occur when differentiating \eqref{eq:ode_q} with respect to $q_0$, and they are functions only of $Re$, and not wavenumbers.

Before turning to calculations,
it worth discussing what these infinitesimal balances equations tell us about neutral modes $(\hat u_0, \hat u_1, \hat v_1, \hat q_0)$. In the case of a homogeneous bifurcation with wavenumbers $k_x = k_z = 0$, (e.g. the saddle-node bifurcation at $Re_{\rm sn}$ in figure~\ref{fig:local_dyn}b), the components $\hat u_0$ and $\hat v_1$ are zero. 
The non-zero components $\hat u_1$ and $\hat q_0$ simultaneously satisfy the momentum balance of the streamwise shear, equation \eqref{eq:hatu1_4} with the left-hand-side equal to zero, and TKE balance \eqref{eq:hatq0_4}. In other words, the ratio of infinitesimals $\hat u_1$ and $\hat q_0$ is such as to neither increase or decrease the streamwise momentum nor the TKE, pointing to a spatially-uniform balances of the standard type that maintain wall-bounded shear turbulence.

For the bifurcation to patterns, the TKE balance now includes turbulent diffusion and advection in equation \eqref{eq:Lop_ktheta}. The momenta balances are, however, now more interesting as they involve many additional effects. 
From equation \eqref{eq:hatv1_4}, for $k_x \ne 0$, the streamwise variation of the Reynolds stress force induces a wall-normal flow $\hat v_1$.
This wall-normal flow produces a momentum source in the streamwise bulk momentum equation \eqref{eq:hatu0_4} via lift up of the base flow $\uss$. The streamwise variation of $\hat u_1$ is also a source of streamwise bulk momentum.
Together, these are balanced by the pressure gradient and dissipation from the streamwise bulk perturbation $\hat u_0$.
Then in the streamwise shear momentum \eqref{eq:hatu1_4}, the Reynolds stress force acts against not only dissipation (as it does for homogenous bifurcations), but additionally against the momentum source from spatial variation of $\hat u_0$. 
All balances hold simultaneously at a bifurcation, but as we now see, it is the advection and pressure terms in the streamwise bulk balance, equation \eqref{eq:hatu0_4}, that are particularly significant for gaining insight into how the band angle is determined.

\subsection{Angle selection}
\label{sec:crit_angle}

From the marginal stability equations, we derive a bound on the band angle at the onset of patterns. All calculations follow exactly equations~\eqref{eq:hat_4}, and no further approximations are made. 
The algebra is straightforward, but we relegate full details to the electronic supplementary material \cite{SM}. 

Using 
\eqref{eq:hatu1_4}, \eqref{eq:hatv1_4}, and \eqref{eq:hatq0_4} to express $\hat{u}_0$, $\hat{u}_1$, and $\hat{q}_0$ in terms of $\hat{v}_1$, one can substitute these into \eqref{eq:hatu0_4} to obtain an equation of the form: $0 = {\rm coefficient} \times \hat{v}_1$. 
Then, since the size of the infinitesimal $\hat{v}_1$ is arbitrary, the coefficient must be zero, giving:
\begin{align}
0 = \underA{
    -\frac{1}{2}\cone(k,\theta,Re) \, k^2 \sin^2 \theta 
    -\frac{1}{2}k^2 \sin^2 \theta
}
    +
\underP{\vphantom{\frac{1}{2}}
    k^2 \sin^4 \theta
}
    + 
\underD{
    \alpha_0 \left(\frac{1 - \cone(k,\theta,Re)}{\ctwo(Re)} \right)
}, 
\label{eq:zero_ev}
\end{align}
where we have defined two combinations of model parameters and closures
\begin{equation}
    \cone(k,\theta,Re) = \frac{2 \myD}{\beta^2 \Ass \dA} \left( \Lzero(Re) - k^2 \Lplus(Re,\theta) \right), \quad 
    \ctwo(Re) = \frac{\uss^2}{\myD}.
\end{equation}
We explicitly indicate the dependencies on $k$, $\theta$, and $Re$. 

Letting $\Fzero(k,\theta,Re)$ denote the right-hand side of \eqref{eq:zero_ev}, the condition for a zero eigenvalue, $\sigma=0$, is written $0 = \Fzero(k,\theta,Re)$.
This is simply the statement that the determinant of the linear stability matrix is zero at a bifurcation, but written with readily identifiable physical meanings for each term. 
We are interested in the critical condition where a zero eigenvalue first appears with decreasing $Re$. 
At such a point we much have 
\begin{subequations}
\label{eq:critical}
\begin{tabular}{p{0.3\textwidth} p{0.3\textwidth} p{0.3\textwidth}}
   {\begin{equation} 
   0 = \Fzero(k_c,\theta_c,Re_c),  \label{eq:critical_ev}
   \end{equation}}
   & 
   {\begin{equation}
   0 = \left .\frac{\partial}{\partial k} \Fzero \right |_{k_c,\theta_c,Re_c}, \label{eq:critical_k}
   \end{equation}}
   &
   {\begin{equation}
   0 = \left .\frac{\partial}{\partial \theta} \Fzero \right |_{k_c,\theta_c,Re_c}, 
    \label{eq:critical_theta}
   \end{equation}}
\end{tabular}
\end{subequations}
where $Re_c$ is the critical Reynolds number and $(k_c, \theta_c$) are the associated critical wavenumber and critical angle 
\footnote{To understand why \eqref{eq:critical_theta} must hold, suppose that $\partial_\theta \Fzero \ne 0$ at the critical point $(k_c, \theta_c, Re_c)$. Then by the implicit function theorem, there is a function $\theta = g(Re)$ such that $\Fzero(k_c, g(Re), Re) = 0$ for $Re$ on an open interval containing $Re_c$. This implies that there is a zero eigenvalue both above and below $Re_c$. The contradicts the assumption that $Re_c$ is the maximum $Re$ for which there is a zero eigenvalue. A similar argument applies fixing $\theta$ and varying $k$.}.
In principle, given the model closure parameters, the three expressions \eqref{eq:critical} could be evaluated and used to find $Re_c$, $k_c$, and $\theta_c$. We have not found this to be analytically viable. However, the condition on the derivative with respect to angle \eqref{eq:critical_theta} itself can be used to establish a bound on $\theta_c$. 

In evaluating \eqref{eq:critical_theta}, that is in differentiating \eqref{eq:zero_ev} with respect to $\theta$, one must take into account that $\cone(k,\theta,Re)$ depends on $\theta$. While this is not difficult, the algebra obscures the essential ideas. For presentation purposes, let us momentarily ignore the $\theta$-dependence of $\cone$. One can then easily differentiate \eqref{eq:zero_ev}, cancel the common factors, and evaluate at $(k_c, \theta_c, Re_c)$ to arrive at
\begin{align}
    0 = -\cone(k_c,\theta_c,Re_c) 
    -1
    +
    4 \sin^2 \theta_c.
\end{align}
From which
\begin{align}
    \sin^2 \theta_c = \frac{1}{4}\left( 1 + \cone(k_c,\theta_c,Re_c)\right).
\label{eq:theta_1}
\end{align}
One sees clearly that this expression arises because the advection and pressure terms in \eqref{eq:zero_ev} have different dependencies on $\theta$. 
Taking into account the $\theta$-dependence of $\Lplus$, the exact statement is instead the inequality (see electronic supplementary material \cite{SM}):
\begin{align}
    \sin^2 \theta_c < \frac{1}{4}\left( 1 + \cone(k_c,\theta_c,Re_c) \right).
\label{eq:theta_2}
\end{align}

We have incorporated all the model closures and parameters into the expression $\cone(k,\theta,Re)$, which we now bound independently of $(k,\theta,Re)$ using a very general argument.
Consider the system's homogeneous dynamics governed by equations~\eqref{eq:odes}. The stability of $(\uss,\qss)$ to uniform perturbations is determined by the eigenvalue problem
\begin{equation}
    \sigma 
    \begin{pmatrix}
         \hat{u}  \\
         \hat{q} 
    \end{pmatrix}
    = 
    \begin{bmatrix}
       -\myD &  - \beta \dA \\
       \frac{1}{2}\beta \Ass & 
       \Lzero
    \end{bmatrix}
    \begin{pmatrix}
         \hat{u}  \\
         \hat{q} 
    \end{pmatrix}.
\end{equation}

For $(\uss,\qss)$ to be a linearly stable steady state of \eqref{eq:odes}, the determinant of the $2 \times 2$ stability matrix must be positive. This gives
\begin{align}
    -\myD \Lzero
    + \frac{1}{2}\beta^2 \Ass \dA & > 0 
\end{align}
or
\begin{align}
    \frac{2 \myD }{\beta^2 \Ass \dA} \Lzero  < 1
\end{align}

Referring back to the definition of $\cone$, we have
\begin{align}
\cone(k,\theta,Re) =  \frac{2 \myD }{\beta^2 \Ass \dA} \left( \Lzero -k^2 \Lplus(\theta,Re) \right)
\le
    \frac{2 \myD }{\beta^2 \Ass \dA} \Lzero  
 < 1.
\end{align}
We have used that all terms in $\Lplus$, equation \eqref{eq:Lop_ktheta}, are positive. Thus we have established $\cone(k,\theta,Re) < 1$, independently of $(k,\theta,Re)$, for any uniform turbulent state that is linearly stable to spatially uniform perturbations. 
Using this bound in \eqref{eq:theta_2} gives $\sin^2 \theta_c < 1/2$. Moreover, evaluating equation \eqref{eq:zero_ev} at $\theta=0$ gives $0 = 1 - \cone(k,0,Re)$, which is impossible since $\cone(k,\theta,Re) < 1$, thus ruling out the possibility of  
any instability to a mode with $\theta=0$. Combining these two conditions, we obtain a bound on the critical angle:
\begin{equation}  
    0 < \sin^2 \theta_c < \frac{1}{2}.
    \label{eq:theta_bound}
\end{equation}
{\em Hence, stability of the uniform turbulent state to uniform perturbations implies a bound on the critical angle for turbulent bands at onset: $ 0 < | \theta_c| < 45^\circ$}. 

While this is a statement about the onset of bands in the model, the result is independent of the particular choices for the turbulence closures. It arises because the advection and pressure terms in the streamwise momentum balance have different dependencies on $\theta$. This follows from basic incompressible hydrodynamics and the result is thus quite general. 
It is possible to improve the bound by exploiting the specific closures used in the model, but we leave this for future work.


\section{Conclusion}
In this work, we have developed a simplified model for transitional turbulence in a planar shear flow. Through Reynolds averaging, vertical-mode truncation, and suitable turbulence closures, we have arrived at a model in six smooth fields
that reproduces many phenomena observed in experiments and simulations, and yet is simple enough to facilitate analysis. We have achieved three distinct advances over the existing literature.

First, we have obtained a model incorporating the large-scale flow associated with transitional turbulence in a planar flow geometry. Advection plays a dominant role in transitional turbulence and the large-scale flow is an essential ingredient to faithful modelling. 
In earlier work on pipe flow \cite{barkley2011simplifying,barkley2015rise,barkley2016theoretical}, a single scalar variable (the mean shear or, equivalently, the centreline velocity) was shown to be sufficient to capture the main effects of the large-scale mean flow. This was possible because, in pipes, flow is confined transverse to the streamwise direction. 
In contrast, planar cases are unconstrained in the spanwise direction and the large-scale flows are correspondingly richer. This has held back development of planar models. 
By considering stress-free boundaries and a sinusoidal driving force, we have succeeded in capturing the incompressible, vector character of the large-scale flow using a minimal expansion in five fields. 

Our second advance is in connecting model equations directly to the Navier-Stokes equations. 
Our model is obtained by projecting the standard Reynolds-averaged Navier-Stokes equations and turbulent-kinetic-energy equation onto a small set of wall-normal modes, and then introducing model closures for the usual higher-order statistical quantities: Reynolds stress, turbulent dissipation and turbulent diffusion. The closures are justified against statistical quantities extracted from direct numerical simulations of the Navier-Stokes equations. 

Our third contribution is new theoretical insights into turbulent bands -- the key building block of transitional turbulence in a planar setting. 
In the model, periodic turbulent bands (turbulent-laminar patterns) unambiguously bifurcate via a linear instability of uniform turbulence as $Re$ is decreased. While this linear instability is expected given the literature on turbulent-laminar patterns \cite{tuckerman2020patterns}, especially recent work by Kashyap {\em et al.} \cite{kashyap2022linear}, the model captures the instability in a simple linear system for momentum and energy balances. 
Analysing this linear system we establish a selection criterion for the pattern orientation at onset, namely, $ 0 < |\theta_c| < 45^\circ$, where the critical angle $\theta_c$ is the angle of the pattern to the streamwise direction when instability first occurs with decreasing $Re$.
Oblique periodic bands in planar shear flows have been ubiquitously observed at angles $45^\circ$ or smaller, often in the range $24^\circ$ to $30^\circ$, and never at $0^\circ$ \cite{tuckerman2020patterns,chantry_universal,prigent2003long,duguet2010formation,duguet2013oblique,ShimizuPRF2019,kashyap2022linear,klotz2022phase,xiong2015turbulent,xiao2020growth, kashyap2020flow, reetz2019exact}. 
Our analysis provides physical insight into what selects such angles -- the relationship between advection and pressure in the streamwise momentum. 
Moreover, the bound suggests a reason why turbulent structures (puffs) in pipes do not arise as periodic patterns from uniform turbulence \cite{moxey2010distinct}. Confinement forces the wavevector to point in the streamwise direction, corresponding to $\theta = 90^\circ$, which is outside the allowed range of angles for instability to periodic patterns.

We offer some further comments and observations on the model. 
We find that the linear instability of spatially uniform turbulence and the existence of bands are extremely robust to changes in closure parameters. Specifically, all the dynamical behaviour illustrated in figure~\ref{fig:panels_1D} is observed over a wide range of closures.
However, the full nonlinear dynamics of turbulent bands, such as depicted in figure~\ref{fig:panels_big}, are more sensitive to model details. 
As discussed in \S \ref{subsec:params}, this sensitivity appears to be primarily due to the reduced dissipation of the large-scale flow from the modal truncation and stress-free boundaries, and is not due to the model closures in the turbulent kinetic energy equation. 
The closures for the Reynolds stress, turbulent dissipation and turbulent diffusion are all based on DNS and standard turbulence theory. We have, however, opted here for mathematical simplicity over precise quantitative agreement with DNS, and this is an area that could be explored further in the future.   

Finally, the model opens several new avenues of research. 
One could add new mechanisms or modify the model to describe other flow configurations, such as pressure-driven pipes and channels, and rotating flows
\cite{meseguer2009instability,berghout2020direct,feldmann2023routes,tsukahara2010flow}. 
One of the more important avenues to pursue is a detailed understanding of the transition scenario using dynamical systems theory, similar to the successful analysis of models of pipe flow \cite{barkley2016theoretical,frishman2022dynamical,frishman2022mechanism}. Closely related to this would be to exploit the physical mechanisms captured by the model to understand how energy balances adjust and break down with decreasing $Re$ \cite{gome2023patterns1,gome2023patterns2}.
As stated in \S\ref{sec:crit_angle}, it should be possible to improve the bound on the angle selection by taking into account details of the turbulent field.
Finally, the effect of turbulent fluctuations can be included via a noise term to investigate rare events \cite{rolland2018extremely,rolland_2022,gome2022extreme} and percolation transitions \cite{lemoult2016directed,chantry_universal,
klotz2022phase,hof2023directed} in the planar setting.


\vskip6pt

\enlargethispage{20pt}

\dataccess{
The code used to run the model simulations described in the main text is v1.0 of `2D MWF' \cite{benavides2025code}. It is preserved at \url{https://doi.org/10.5281/zenodo.15261063}, available via the Creative Commons Attribution 4.0 International (CC-BY) license. The code depends largely on the Dedalus package in Python \cite{burns2020dedalus}, an open-source pseudo-spectral solver for partial differential equations.
The electronic supplementary material \cite{SM}, preserved in the Figshare repository \url{https://doi.org/10.6084/m9.figshare.28903058}, contains (i) details of the stability analysis, bound on the critical angle and description of the supplementary movies, (ii) the supplementary movies, (iii) the data used for the figures in this work, and (iv) the scripts used to generate the figures in this work.
} 

\AI{We have not used AI-assisted technologies in creating this article.}

\aucontribute{S.J.B.: conceptualization, formal analysis, investigation, methodology, visualization, writing -- original draft, writing -- review and editing; D. B. : conceptualization, formal analysis, funding acquisition, investigation, methodology, visualization, writing -- original draft, writing -- review and editing.}

\conflict{We declare we have no competing interests.}

\funding{This work was supported by a grant from the Simons Foundation (grant no. 662985).}

\ack{We thank S{\'e}bastien Gom{\'e} and Yohann Duguet, and Laurette Tuckerman for feedback and valuable discussions.}

\appendix

\section{Model equation in a 1D tilted domain \label{app:tilted}}

In this appendix we present the model equations restricted to solutions at a prescribed angle $\theta$ with respect to the streamwise direction. 
For this purpose, we work in horizontal coordinates $(\xrot,\zrot)$ rotated by angle $\theta$ with respect to the streamwise-spanwise coordinates $(x,z)$.
The bases vectors, coordinates, and horizontal velocity components obey
\begin{align}
\exrot & = \cos \theta \, \ex - \sin \theta \, \ez,
& \ezrot = \sin \theta \, \ex + \cos \theta \, \ez \\
\xrot & = \cos \theta \, x + \sin \theta \, z,
& \zrot = -\sin \theta x + \cos \theta \, z, \\
\urot & = \cos \theta \, u - \sin \theta \, w,
& \wrot = \sin \theta \, u + \cos \theta \, w,
\end{align}
where rotated quantities appear on the left. We take $\exrot$ to point in the along-band direction and $\ezrot$ in the across-band direction. The wall-normal coordinate $y$ and velocity $v$, and scalar TKE $q$ are unaffected by horizontal rotation. 
The model simplification comes by considering solutions that are independent of the along-band direction $\xrot$. Hence, model fields become functions of $(\zrot, t)$ only and all along-band derivatives, $\partial_{\xrot}$, vanish from the governing equations.

While applying the change of coordinates is straightforward, the system is not isotropic and some terms in the governing equations explicitly depend on streamwise direction $\ex$. These terms pick up $\cos \theta$ or $\sin \theta$ factors when expressed in rotated coordinates. Specifically, the driving body force takes the form
\begin{equation*}
\fw = (\alpha + \beta^2/Re) \sin\left(\beta y\right) \ex 
= (\alpha + \beta^2/Re) \sin\left(\beta y\right) (\cos \theta \, \exrot + \sin \theta \, \ezrot)
\end{equation*}
The divergence of the Reynolds stress takes the form
\begin{align*}
\nabla \cdot \mathcal{R} & = - \beta A(q_0) \sin(\beta y) \ex + \partial_x A(q_0) \cos(\beta y) \ey \\
& = - \beta A(q_0) \sin(\beta y) (\cos \theta \, \exrot + \sin \theta \, \ezrot) + \sin \theta \, \partial_\zrot A(q_0) \cos(\beta y) \ey,
\end{align*}
where in the last term we have used that $\partial_x = \cos \theta \, \partial_{\xrot} + \sin \theta  \, \partial_{\zrot} = \sin \theta \, \partial_{\zrot}$. Finally, the production terms take the form
\begin{align*}
\mathcal{P} = (\partial_y u + \partial_x v) A(q_0) \cos(\beta y) = (\cos \theta \, \partial_y \urot + \sin \theta \, \partial_y \wrot + \sin \theta \,\partial_\zrot v) A(q_0) \cos(\beta y)
\end{align*}

Thus, the model equations in rotated coordinates become
\begin{equation*}
\begin{aligned}
    \frac{\partial \urot_0}{\partial t}  &+ \frac{1}{2} (\beta v_1  + \wrot_1 \partial_{\zrot}) \urot_1 
    = \left(\frac{1}{Re}\partial_{\zrot}^2 - \alpha_0 \right) \urot_0,  \\
    \frac{\partial \urot_1}{\partial t} & + \wrot_1 \partial_{\zrot} \urot_0 
    = \left(\frac{1}{Re}(\partial_{\zrot}^2 - \beta^2) - \alpha \right) \urot_1 +
   f \cos\theta
    - \beta A(q_0) \cos\theta, \\    
    \frac{\partial v_1}{\partial t} 
    &=  -\beta p_1 + \frac{1}{Re}(\partial_{\zrot}^2 - \beta^2) v_1 - \partial_{\zrot} A(q_0) \sin \theta, \\
    \frac{\partial \wrot_1}{\partial t} & = -\partial_{\zrot} p_1 + \left(\frac{1}{Re}(\partial_{\zrot}^2 - \beta^2) - \alpha \right) \wrot_1 + \beta A(q_0) \sin\theta - f \sin\theta, \\    
    \frac{\partial q_0}{\partial t} &+ \frac{1}{2} (\beta v_1 + \wrot_1 \partial_{\zrot}) q_1(\urot_1,\wrot_1,q_0)  = \frac{1}{2}A(q_0) [\beta (\urot_1 \cos\theta - \wrot_1 \sin\theta) - \partial_{\zrot} v_1 \sin\theta] \\ & -2\alpha q_0 - \varepsilon_0(q_0;Re) + \frac{1}{Re}\partial_{\zrot}^2 q_0 + \partial_{\zrot} \left(\nu_T(q_0;Re) \partial_{\zrot} q_0 \right),
\end{aligned}
\label{eq:model_1D}
\end{equation*}
where $f=\alpha + \beta^2/Re$ and $\beta = \pi/2$. The fields $\urot_0(\zrot,t)$ and $\urot_1(\zrot,t)$ capture flow along the bands, while $\wrot_1(\zrot,t)$ captures flow across the bands. The field $\wrot_0$, describing the vertically constant across-band flow, vanishes by incompressibility: $\partial_{\xrot} \urot_0 + \partial_{\zrot} \wrot_0 = \partial_{\zrot} \wrot_0 = 0$, from which $\wrot_0 = 0$, assuming no bulk flow. 
The system of equations can be further reduced to only four independent fields, since incompressibility of the first vertical modes implies $\beta v_1 = \partial_{\zrot} \wrot_1$. Two possible approaches are: replace the evolution equations for the $v_1$ and $\wrot_1$ fields with a single evolution equation for the vorticity component $\zeta_1 = \beta \wrot_1 - \partial_\zrot v_1$, or use incompressibility to obtain an expression for the pressure $p_1$. The two approaches are ultimately equivalent and result in the simplest formulation of the 1D model equations.
We have solved these equations numerically to produce figure~\ref{fig:panels_1D}.

\section{Linear-stability equations \label{app:lin_stab}}

In this appendix we provide a complete statement of the linear-stability equations. Uniform steady states are of the form $(u_0, u_1, v_1, w_0, w_1, q_0) = (0,\uss,0,0,0,\qss)$. Straightforward linearization of the full model about such a steady state gives the following eigenvalue problem for the temporal growth rate $\sigma$ of a linear mode, denoted with hats, with wavenumbers $(k_x, k_z)$
\begin{equation}
\begin{aligned}
& \sigma \hat{u}_0 + \frac{1}{2} \uss (i k_x \hat{u}_1 + \beta \hat{v}_1) = \frac{k_x^2 \beta \uss }{k^2} \, \hat{v}_1 - \left({\frac{k^2}{Re}} + \alpha_0 \right) \hat{u}_0,  \\
& \sigma \hat{u}_1 + i k_x \uss \hat{u}_0 = 
-{\frac{ik_x \alpha \beta}{k^2+\beta^2}} \, \hat{v}_1
- \left(\frac{{k^2} + \beta^2}{Re} + \alpha \right) \hat{u}_1 - \left(1 - {\frac{2k_x^2}{k^2+\beta^2}}\right) \beta \dA 
 \, \hat{q}_0, \\
& \sigma \hat{v}_1 = - \alpha {\frac{\beta^2}{k^2+\beta^2}} \hat{v}_1 -\frac{{k^2} + \beta^2}{Re} \hat{v}_1 + \left(1 - {\frac{2\beta^2}{k^2+\beta^2}}\right) ik_x \dA \hat{q}_0, \\
& \sigma \hat{w}_0 + \frac{1}{2} ik_x \uss \hat{w}_1 =  \frac{k_x k_z \beta \uss}{k^2} \, \hat{v}_1
 - \left({\frac{k^2}{Re}} + \alpha_0\right) \hat{w}_0,  \\
& \sigma \hat{w}_1 + ik_x \uss \hat{w}_0 =  
-{\frac{ik_z \alpha \beta}{k^2+\beta^2}} \, \hat{v}_1
- \left(\frac{{k^2} + \beta^2}{Re} + \alpha \right) \hat{w}_1 +{\frac{2k_x k_z \beta \dA }{k^2+\beta^2} }
 \, \hat{q}_0, \\
& \sigma \hat{q}_0 = \frac{1}{2} \beta \Ass  \hat{u}_1 + {\frac{1}{2} \Ass ik_x} \hat{v}_1  + \Lop \hat{q}_0,
\end{aligned}
\label{eq:lse}
\end{equation}
where $k^2 = k_x^2 + k_z^2$, and

\begin{equation}
    \Ass \equiv A(\qss), \quad
    \dA \equiv \frac{\partial A}{\partial q_0}(\qss), \quad \deps \equiv \frac{d \varepsilon_0}{d q_0}(\qss,Re),
\end{equation}

\begin{equation}
 \Lop  = 
    \frac{1}{2}\beta \uss \dA 
    - 2 \alpha 
    - \deps 
    - \frac{k^2}{Re} 
    - \nu_T(\qss;Re) k^2
    - \frac{\uss^2 k_x^2}{2(2 \alpha + \kappa + \beta^2/Re)}.
\label{eq:Lop_hat_foo}
\end{equation}
 
In the long-wavelength approximation, $k/\beta \ll 1$, many terms can be neglected and the linear-stability equations simplify to
\begin{equation}
\begin{aligned}
& \sigma \hat{u}_0 + \frac{1}{2} \uss (i k_x \hat{u}_1 + \beta \hat{v}_1) = \beta \uss \frac{k_x^2 }{k^2} \, \hat{v}_1 - \alpha_0 \hat{u}_0,  \\
& \sigma \hat{u}_1 + i k_x \uss \hat{u}_0 = 
- \left(\alpha + \beta^2/Re \right) \hat{u}_1 - \beta \dA 
 \, \hat{q}_0, \\
& \sigma \hat{v}_1 = - \left(\alpha + \beta^2/Re \right) \hat{v}_1 - ik_x \dA \hat{q}_0, \\
& \sigma \hat{w}_0 + \frac{1}{2} ik_x \uss \hat{w}_1 = \beta \uss \frac{k_x k_z }{k^2} \, \hat{v}_1
 - \alpha_0 \hat{w}_0,  \\
& \sigma \hat{w}_1 + ik_x \uss \hat{w}_0 =  
- \left(\alpha + \beta^2/Re \right) \hat{w}_1, \\
& \sigma \hat{q}_0 = \frac{1}{2} \beta \Ass  \hat{u}_1 + \Lop \hat{q}_0.
\end{aligned}
\label{eq:lse_long}
\end{equation}
For a detailed derivation of the linear-stability equations, see the electronic supplementary material \cite{SM}.


\vskip2pc

\bibliographystyle{RS}
\bibliography{Minimal}

@article{tsukahara2010flow,
  title={Flow regimes in a plane Couette flow with system rotation},
  author={Tsukahara, Takahiro and Tillmark, Nils and Alfredsson, PH},
  journal={J. Fluid Mech.},
  volume={648},
  pages={5--33},
  year={2010}
}

@article{feldmann2023routes,
  title={Routes to turbulence in Taylor--Couette flow},
  author={Feldmann, Daniel and Borrero-Echeverry, Daniel and Burin, Michael J and Avila, Kerstin and Avila, Marc},
  journal={Phil. Trans. R. Soc. A},
  volume={381},
  number={2246},
  pages={20220114},
  year={2023},
  publisher={The Royal Society}
}

@ARTICLE{Farrell2017,
       author = {{Farrell}, B.~F. and {Gayme}, D.~F. and {Ioannou}, P.~J.},
        title = "{A statistical state dynamics approach to wall turbulence}",
      journal = {Phil. Trans. R. Soc. A},
         year = 2017,
        month = mar,
       volume = {375},
       number = {2089},
          eid = {20160081},
        pages = {20160081},
}

@article{lemoult2024directed,
  title={Directed percolation and puff jamming near the transition to pipe turbulence},
  author={Lemoult, Gr{\'e}goire and Mukund, Vasudevan and Shih, Hong-Yan and Linga, Gaute and Mathiesen, Joachim and Goldenfeld, Nigel and Hof, Bj{\"o}rn},
  journal={Nature Physics},
  volume={20},
  number={8},
  pages={1339--1345},
  year={2024},
  publisher={Nature Publishing Group UK London}
}

@article{kashyap2025laminar,
  title={Laminar-Turbulent Patterns in Shear Flows: Evasion of Tipping, Saddle-Loop Bifurcation, and Log Scaling of the Turbulent Fraction},
  author={Kashyap, Pavan V and Mar{\'\i}n, Juan F and Duguet, Yohann and Dauchot, Olivier},
  journal={Physical Review Letters},
  volume={134},
  number={15},
  pages={154001},
  year={2025},
  publisher={APS}
}

@article{tollmien1935,
  title={Ein allgemeines Kriterium der Instabilit{\"a}t laminarer Geschwindigkeitsverteilung},
  author={Tollmien, Walter},
  journal={Nachr. Ges. Wiss. G{\"o}ttingen, Math.-phys. Klasse},
  volume={1},
  pages={79--114},
  year={1935}
}

@article{manneville2017laminar,
  title={Laminar-turbulent patterning in transitional flows},
  author={Manneville, Paul},
  journal={Entropy},
  volume={19},
  number={7},
  pages={316},
  year={2017},
  publisher={MDPI}
}

@article{philip2011temporal,
  title={From temporal to spatiotemporal dynamics in transitional plane Couette flow},
  author={Philip, Jimmy and Manneville, Paul},
  journal={Phys.~Rev.~E},
  volume={83},
  number={3},
  pages={036308},
  year={2011},
  publisher={APS}
}

@article{wang2022stochastic,
  title={Stochastic model for quasi-one-dimensional transitional turbulence with streamwise shear interactions},
  author={Wang, Xueying and Shih, Hong-Yan and Goldenfeld, Nigel},
  journal={Phys.~Rev.~Lett.},
  volume={129},
  number={3},
  pages={034501},
  year={2022},
  publisher={APS}
}

@article{avila2023transition,
  title={Transition to turbulence in pipe flow},
  author={Avila, Marc and Barkley, Dwight and Hof, Bj{\"o}rn},
  journal={Annu.~Rev.~Fluid Mech.},
  volume={55},
  pages={575--602},
  year={2023},
  publisher={Annual Reviews}
}

@article{manneville2015transition,
  title={On the transition to turbulence of wall-bounded flows in general, and plane Couette flow in particular},
  author={Manneville, Paul},
  journal={Eur. J. Mech. B/Fluids},
  volume={49},
  pages={345--362},
  year={2015},
  publisher={Elsevier}
}

@article{manneville2016review,
  title={Transition to turbulence in wall-bounded flows: Where do we stand?},
  author={Paul Manneville},
  journal={Mech. Eng. Rev.},
  volume={3},
  number={2},
  pages={15-00684-15-00684},
  year={2016},
}

@article{kashyap2022linear,
  title={Linear instability of turbulent channel flow},
  author={Kashyap, Pavan V and Duguet, Yohann and Dauchot, Olivier},
  journal={Phys.~Rev.~Lett.},
  volume={129},
  number={24},
  pages={244501},
  year={2022},
  publisher={APS}
}

@article{gome2023patterns1,
  title={Patterns in transitional shear turbulence. Part 1. Energy transfer and mean-flow interaction},
  author={Gom{\'e}, S{\'e}bastien and Tuckerman, Laurette S and Barkley, Dwight},
  journal={J.~Fluid Mech.},
  volume={964},
  pages={A16},
  year={2023},
  publisher={Cambridge University Press}
}

@article{gome2023patterns2,
  title={Patterns in transitional shear turbulence. Part 2. Emergence and optimal wavelength},
  author={Gom{\'e}, S{\'e}bastien and Tuckerman, Laurette S and Barkley, Dwight},
  journal={J.~Fluid Mech.},
  volume={964},
  pages={A17},
  year={2023},
  publisher={Cambridge University Press}
}

@article{lundbladh1991direct,
  title={Direct simulation of turbulent spots in plane Couette flow},
  author={Lundbladh, Anders and Johansson, Arne V},
  journal={J.~Fluid Mech.},
  volume={229},
  pages={499-516},
  year={1991}
}

@article{frishman2022dynamical,
  title={Dynamical landscape of transitional pipe flow},
  author={Frishman, Anna and Grafke, Tobias},
  journal={Phys.~Rev.~E},
  volume={105},
  number={4},
  pages={045108},
  year={2022},
  publisher={APS}
}

@article{frishman2022mechanism,
  title={Mechanism for turbulence proliferation in subcritical flows},
  author={Frishman, Anna and Grafke, Tobias},
  journal={Proc. Roy. Soc. A},
  volume={478},
  number={2265},
  pages={20220218},
  year={2022},
  publisher={The Royal Society}
}

@article{manneville2011modelling,
  title={On modelling transitional turbulent flows using under-resolved direct numerical simulations: the case of plane Couette flow},
  author={Manneville, Paul and Rolland, Joran},
  journal={Theor. Comput. Fluid Dyn.},
  volume={25},
  pages={407--420},
  year={2011},
  publisher={Springer}
}

@article{seshasayanan2015laminar,
  title={Laminar-turbulent patterning in wall-bounded shear flows: a Galerkin model},
  author={Seshasayanan, K and Manneville, Paul},
  journal={Fluid Dyn. Res.},
  volume={47},
  number={3},
  pages={035512},
  year={2015},
  publisher={IOP Publishing}
}

@article{marcus1998model,
  title={A model for eastward and westward jets in laboratory experiments and planetary atmospheres},
  author={Marcus, PS and Lee, C},
  journal={Physics of Fluids},
  volume={10},
  number={6},
  pages={1474--1489},
  year={1998},
  publisher={American Institute of Physics}
}

@book{pedlosky2013geophysical,
  title={Geophysical fluid dynamics},
  author={Pedlosky, Joseph},
  year={2013},
  publisher={Springer Science \& Business Media}
}

@article{lu2022growth,
  title={Growth and decay of isolated turbulent band in Plane \protect{Couette} flow}, 
  author={Jianzhou Lu and Jianjun Tao and Weitao Zhou},
  year={2022},
  journal={arXiv:2304.12409},
}

@article{manneville2012growthspot,
year = {2012},
month = {may},
publisher = {IOP Publishing},
volume = {44},
number = {3},
pages = {031412},
author = {Paul Manneville},
title = {On the growth of laminar–turbulent patterns in plane Couette flow},
journal = {Fluid Dyn.~Res.},
}

@article{schumacher2001largescaleflow,
  title = {Evolution of turbulent spots in a parallel shear flow},
  author = {Schumacher, J\"org and Eckhardt, Bruno},
  journal = {Phys. Rev. E},
  volume = {63},
  issue = {4},
  pages = {046307},
  numpages = {9},
  year = {2001},
  month = {Mar},
  publisher = {American Physical Society},
}

@article{lagha2007largescaleflow,
    author = {Lagha, Maher and Manneville, Paul},
    title = "{Modeling of plane Couette flow. I. Large scale flow around turbulent spots}",
    journal = {Phys. Fluids},
    volume = {19},
    number = {9},
    pages = {094105},
    year = {2007},
    month = {09}
}

@article{lagha2007modeling,
  title={Modeling transitional plane Couette flow},
  author={Lagha, Maher and Manneville, Paul},
  journal={The European Physical Journal B},
  volume={58},
  pages={433--447},
  year={2007},
  publisher={Springer}
}

@article{wang2020quadrupolar, title={Quadrupolar flows around spots in internal shear flows}, volume={892}, journal={J.~Fluid Mech.}, publisher={Cambridge University Press}, author={Wang, Zhe and Guet, Claude and Monchaux, Romain and Duguet, Yohann and Eckhardt, Bruno}, year={2020}, pages={A27}}

@software{benavides2025code,
  author       = {Santiago J Benavides},
  title        = {s-benavides/2D\_MWF: 2D Model Waleffe Flow v1.0},
  month        = apr,
  year         = 2025,
  publisher    = {Zenodo},
  version      = {v1.0},
  doi          = {10.5281/zenodo.15261063},
  url          = {https://doi.org/10.5281/zenodo.15261063},
  swhid        = {swh:1:dir:d34b01908df0284d3d733b0a54de0546b52e1da5
                   ;origin=https://doi.org/10.5281/zenodo.15261062;vi
                   sit=swh:1:snp:f44cc671b70c1b95a93ed6461f216673dc58
                   dc43;anchor=swh:1:rel:8032ee0181ff37e956da6460ee85
                   125b74e99d32;path=s-benavides-2D\_MWF-788330f
                  },
}

@article{burns2020dedalus,
  title = {Dedalus: A flexible framework for numerical simulations with spectral methods},
  author = {Burns, Keaton J. and Vasil, Geoffrey M. and Oishi, Jeffrey S. and Lecoanet, Daniel and Brown, Benjamin P.},
  journal = {Phys. Rev. Res.},
  volume = {2},
  issue = {2},
  pages = {023068},
  numpages = {39},
  year = {2020},
  month = {Apr},
  publisher = {American Physical Society},
}

@article{suri2014drag,
    author = {Suri, Balachandra and Tithof, Jeffrey and Mitchell, Radford, Jr. and Grigoriev, Roman O. and Schatz, Michael F.},
    title = "{Velocity profile in a two-layer Kolmogorov-like flow}",
    journal = {Phys. Fluids},
    volume = {26},
    number = {5},
    pages = {053601},
    year = {2014},
    month = {05},
}

@article{manneville2012turing,
year = {2012},
month = {jun},
publisher = {},
volume = {98},
number = {6},
pages = {64001},
author = {Paul Manneville},
title = {Turbulent patterns in wall-bounded flows: A Turing instability?},
journal = {Europhys. Lett.}
}

@article{prigent2003long,
  title={Long-wavelength modulation of turbulent shear flows},
  author={Prigent, Arnaud and Gr{\'e}goire, Guillaume and Chat{\'e}, Hugues and Dauchot, Olivier},
  journal={Physica D},
  volume={174},
  number={1-4},
  pages={100--113},
  year={2003},
  publisher={Elsevier}
}

@article{hamilton1995regeneration,
  title={Regeneration mechanisms of near-wall turbulence structures},
  author={Hamilton, James M and Kim, John and Waleffe, Fabian},
  journal={J.~Fluid Mech.},
  volume={287},
  pages={317--348},
  year={1995},
  publisher={Cambridge University Press}
}

@article{chantry2016turbulent,
  title={Turbulent--laminar patterns in shear flows without walls},
  author={Chantry, Matthew and Tuckerman, Laurette S and Barkley, Dwight},
  journal={J.~Fluid Mech.},
  volume={791},
  pages={R8},
  year={2016},
  publisher={Cambridge University Press}
}

@article{chantry_universal,
  title={Universal continuous transition to turbulence in a planar shear flow},
  author={Chantry, Matthew and Tuckerman, Laurette S and Barkley, Dwight},
  journal={J.~Fluid Mech.},
  volume={824},
  pages={R1},
  year={2017},
  publisher={Cambridge University Press}
}

@article{waleffe1997self,
  title={On a self-sustaining process in shear flows},
  author={Waleffe, Fabian},
  journal={Phys.~Fluids},
  volume={9},
  number={4},
  pages={883--900},
  year={1997},
  publisher={AIP}
}

@article{barkley2005computational,
  title={Computational study of turbulent-laminar patterns in {Couette} flow},
  author={Barkley, Dwight and Tuckerman, Laurette S},
  journal={Phys.~Rev.~Lett.},
  volume={94},
  number={1},
  pages={014502},
  year={2005},
  publisher={APS}
}

@article{barkley2007mean,
  title={Mean flow of turbulent--laminar patterns in plane {Couette} flow},
  author={Barkley, Dwight and Tuckerman, Laurette S},
  journal={J.~Fluid Mech.},
  volume={576},
  pages={109--137},
  year={2007},
  publisher={Cambridge University Press}
}

@article{tuckerman2011patterns,
  title={Patterns and dynamics in transitional plane {Couette} flow},
  author={Tuckerman, Laurette S and Barkley, Dwight},
  journal={Phys.~Fluids},
  volume={23},
  number={4},
  pages={041301},
  year={2011},
  publisher={AIP}
}

@article{tuckerman2014turbulent,
  title={Turbulent-laminar patterns in plane {Poiseuille} flow},
  author={Tuckerman, Laurette S and Kreilos, Tobias and Schrobsdorff, Hecke and Schneider, Tobias M and Gibson, John F},
  journal={Phys.~Fluids},
  volume={26},
  number={11},
  pages={114103},
  year={2014},
  publisher={AIP}
}

@article{darbyshire1995transition,
  title={Transition to turbulence in constant-mass-flux pipe flow},
  author={Darbyshire, AG and Mullin, T},
  journal={J.~Fluid Mech.},
  volume={289},
  pages={83--114},
  year={1995},
  publisher={Cambridge University Press}
}

@article{duguet2010formation,
  title={Formation of turbulent patterns near the onset of transition in plane {Couette} flow},
  author={Duguet, Yohann and Schlatter, Philipp and Henningson, Dan S},
  journal={J.~Fluid Mech.},
  volume={650},
  pages={119--129},
  year={2010},
  publisher={Cambridge University Press}
}

@article{ShimizuPRF2019,
  title = {Bifurcations to turbulence in transitional channel flow},
  author = {Shimizu, Masaki and Manneville, Paul},
  journal = {Phys.~Rev.~Fluids},
  volume = {4},
  issue = {11},
  pages = {113903},
  numpages = {21},
  year = {2019},
  publisher = {American Physical Society},
}

@article{lemoult2016directed,
  title={Directed percolation phase transition to sustained turbulence in {Couette} flow},
  author={Lemoult, Gr{\'e}goire and Shi, Liang and Avila, Kerstin and Jalikop, Shreyas V and Avila, Marc and Hof, Bj{\"o}rn},
  journal={Nat. Phys.},
  volume={12},
  number={3},
  pages={254},
  year={2016},
  publisher={Nature Publishing Group}
}

@article{shih2016ecological,
  title={Ecological collapse and the emergence of travelling waves at the onset of shear turbulence},
  author={Shih, Hong-Yan and Hsieh, Tsung-Lin and Goldenfeld, Nigel},
  journal={Nat. Phys.},
  volume={12},
  number={3},
  pages={245--248},
  year={2016},
  publisher={Nature Publishing Group}
}

@article{tuckerman2020patterns,
  title={Patterns in wall-bounded shear flows},
  author={Tuckerman, Laurette S and Chantry, Matthew and Barkley, Dwight},
  journal={Annu.~Rev.~Fluid Mech.},
  volume={52},
  pages={343},
  year={2020},
  publisher={Annual Reviews}
}

@article{xiao2020growth,
  title={The growth mechanism of turbulent bands in channel flow at low {Reynolds} numbers},
  author={Xiao, Xiangkai and Song, Baofang},
  journal={J.~Fluid Mech.},
  pages={R1},
  volume={883},
  year={2020},
  publisher={Cambridge University Press}
}

@article{xiong2015turbulent,
    author = {Xiong, Xiangming and Tao, Jianjun and Chen, Shiyi and Brandt, Luca},
    title={Turbulent bands in plane-\protect{Poiseuille} flow at moderate \protect{Reynolds} numbers},
    journal = {Physics of Fluids},
    volume = {27},
    number = {4},
    pages = {041702},
    year = {2015},
    month = {04},
}

@article{barkley2016theoretical,
  title={Theoretical perspective on the route to turbulence in a pipe},
  author={Barkley, Dwight},
  journal={J.~Fluid Mech.},
  volume={803},
  pages={P1},
  year={2016},
  publisher={Cambridge University Press}
}

@article{reetz2019exact,
  title={Exact invariant solution reveals the origin of self-organized oblique turbulent-laminar stripes},
  author={Reetz, Florian and Kreilos, Tobias and Schneider, Tobias M},
  journal={Nat. Commun.},
  volume={10},
  number={1},
  pages={2277},
  year={2019},
  publisher={Nature Publishing Group UK London}
}

@article{coles1966progress,
  title={Progress report on a digital experiment in spiral turbulence},
  author={Coles, Donald and van Atta, Charles},
  journal={AIAA Journal},
  volume={4},
  number={11},
  pages={1969--1971},
  year={1966}
}

@article{prigent2002large,
  title={Large-scale finite-wavelength modulation within turbulent shear flows},
  author={Prigent, Arnaud and Gr{\'e}goire, Guillaume and Chat{\'e}, Hugues and Dauchot, Olivier and van Saarloos, Wim},
  journal={Phys.~Rev.~Lett.},
  volume={89},
  number={1},
  pages={014501},
  year={2002},
  publisher={APS}
}

@article{meseguer2009instability,
  title={Instability mechanisms and transition scenarios of spiral turbulence in {Taylor}-{Couette} flow},
  author={Meseguer, Alvaro and Mellibovsky, Fernando and Avila, Marc and Marques, Francisco},
  journal=PRE,
  volume={80},
  number={4},
  pages={046315},
  year={2009},
  publisher={APS}
}

@article{berghout2020direct,
  title={Direct numerical simulations of spiral {Taylor}-{Couette} turbulence},
  author={Berghout, Pieter and Dingemans, Rick J and Zhu, Xiaojue and Verzicco, Roberto and Stevens, Richard JAM and van Saarloos, Wim and Lohse, Detlef},
  journal={J.~Fluid Mech.},
  volume={887},
  year={2020},
  publisher={Cambridge University Press}
}

@article{barkley2011simplifying,
  title={Simplifying the complexity of pipe flow},
  author={Barkley, Dwight},
  journal={Phys. Rev. E},
  volume={84},
  number={1},
  pages={016309},
  year={2011},
  publisher={APS}
}

@article{barkley2015rise,
  title={The rise of fully turbulent flow},
  author={Barkley, Dwight and Song, Baofang and Mukund, Vasudevan and Lemoult, Gr{\'e}goire and Avila, Marc and Hof, Bj{\"o}rn},
  journal={Nature},
  volume={526},
  number={7574},
  pages={550--553},
  year={2015},
  publisher={Nature Publishing Group}
}

@article{song2017speed,
  title={Speed and structure of turbulent fronts in pipe flow},
  author={Song, Baofang and Barkley, Dwight and Hof, Bj{\"o}rn and Avila, Marc},
  journal={J.~Fluid Mech.},
  volume={813},
  pages={1045--1059},
  year={2017},
  publisher={Cambridge University Press}
}

@article{klotz2021experimental,
  title={Experimental measurements in plane {Couette}--{Poiseuille} flow: dynamics of the large-and small-scale flow},
  author={Klotz, Lukasz and Pavlenko, A.M. and Wesfreid, J.E.},
  journal={J.~Fluid Mech.},
  volume={912},
  year={2021},
  publisher={Cambridge University Press}
}

@article{rolland2018extremely,
  title={Extremely rare collapse and build-up of turbulence in stochastic models of transitional wall flows},
  author={Rolland, Joran},
  journal={Phys. Rev. E},
  volume={97},
  number={2},
  pages={023109},
  year={2018},
  publisher={APS}
}

@article{rolland_2022, title={Collapse of transitional wall turbulence captured using a rare events algorithm}, volume={931}, journal={J. Fluid Mech.}, publisher={Cambridge University Press}, author={Rolland, Joran}, year={2022}, pages={A22}}

@article{moxey2010distinct,
  title={Distinct large-scale turbulent-laminar states in transitional pipe flow},
  author={Moxey, David and Barkley, Dwight},
  journal={Proc.~Natl.~Acad.~Sci.~U.S.A.},
  volume={107},
  number={18},
  pages={8091--8096},
  year={2010},
  publisher={National Acad Sciences}
}

@book{pope2000turbulent,
  title={Turbulent flows},
  author={Pope, Stephen B},
  year={2000},
  publisher={\protect{Camb. Univ. Press}}
}

@article{wygnanski1973transition,
  title={On transition in a pipe. Part 1. The origin of puffs and slugs and the flow in a turbulent slug},
  author={Wygnanski, Israel J and Champagne, FH},
  journal={J.~Fluid Mech.},
  volume={59},
  number={2},
  pages={281--335},
  year={1973},
  publisher={Cambridge University Press}
}

@article{kashyap2020flow,
  title={Flow statistics in the transitional regime of plane channel flow},
  author={Kashyap, Pavan V and Duguet, Yohann and Dauchot, Olivier},
  journal={Entropy},
  volume={22},
  number={9},
  pages={1001},
  year={2020},
  publisher={Multidisciplinary Digital Publishing Institute}
}

@article{klotz2022phase,
  title={Phase Transition to Turbulence in Spatially Extended Shear Flows},
  author={Klotz, Lukasz and Lemoult, Gr{\'e}goire and Avila, Kerstin and Hof, Bj{\"o}rn},
  journal={Phys.~Rev.~Lett.},
  volume={128},
  number={1},
  pages={014502},
  year={2022},
  publisher={APS}
}

@article{kashyap2020far,
  title={Far field of turbulent spots},
  author={Kashyap, Pavan V and Duguet, Yohann and Chantry, Matthew},
  journal={Phys.~Rev.~Fluids},
  volume={5},
  number={10},
  pages={103902},
  year={2020},
  publisher={APS}
}

@article{duguet2013oblique,
  title={Oblique laminar-turbulent interfaces in plane shear flows},
  author={Duguet, Yohann and Schlatter, Philipp},
  journal={Phys.~Rev.~Lett.},
  volume={110},
  number={3},
  pages={034502},
  year={2013},
  publisher={APS}
}

@article{gome2022extreme,
  title={Extreme events in transitional turbulence},
  author={Gom{\'e}, S{\'e}bastien and Tuckerman, Laurette S and Barkley, Dwight},
  journal={Philos.~Trans.~R.~Soc.~A},
  volume={380},
  number={2226},
  pages={20210036},
  year={2022},
  publisher={The Royal Society}
}

@article{pomeau2015transition,
  title={The transition to turbulence in parallel flows: a personal view},
  author={Pomeau, Yves},
  journal={Comptes Rendus M{\'e}canique},
  volume={343},
  number={3},
  pages={210--218},
  year={2015},
  publisher={Elsevier}
}

@article{couliou2016spreading,
  title={Spreading of turbulence in plane {Couette} flow},
  author={Couliou, Marie and Monchaux, Romain},
  journal={Phys.~Rev.~E},
  volume={93},
  number={1},
  pages={013108},
  year={2016},
  publisher={APS}
}

@misc{SM,
author = {Benavides, Santiago J. and Barkley, Dwight},
year = {2025},
title = {Supplementary material from: Model for transitional turbulence in a planar shear flow},
note = "Figshare.", 
doi= {10.6084/m9.figshare.28903058},
url= {https://doi.org/10.6084/m9.figshare.28903058},
}

@article{marensi2022dynamics, title={Dynamics and proliferation of turbulent stripes in plane-Poiseuille and plane-Couette flows}, volume={974}, journal={Journal of Fluid Mechanics}, author={Marensi, E. and Yalnız, G. and Hof, B.}, year={2023}, pages={A21}}

@article{hof2023directed,
  title={Directed percolation and the transition to turbulence},
  author={Hof, Bj{\"o}rn},
  journal={Nat. Rev. Phys.},
  volume={5},
  number={1},
  pages={62--72},
  year={2023},
  publisher={Nature Publishing Group UK London}
}

@article{Ohnishi_2011,
year = {2011},
month = {dec},
publisher = {},
volume = {318},
number = {3},
pages = {032033},
author = {Kyohei Ohnishi and Takahiro Tsukahara and Yasuo Kawaguchi},
title = {Turbulence budget in transitional plane Couette flow with turbulent stripe},
journal = {Journal of Physics: Conference Series}
}

@InProceedings{Minnick2024,
author="Minnick, B. A.
and Viggiano, B.
and Gayme, D. F.",
editor="{\"O}rl{\"u}, Ramis
and Talamelli, Alessandro
and Peinke, Joachim
and Oberlack, Martin",
title="Augmented Restricted Nonlinear (ARNL) Model for High Reynolds Number Wall-Turbulence",
booktitle="Progress in Turbulence X",
year="2024",
publisher="Springer Nature Switzerland",
address="Cham",
pages="65--75",
isbn="978-3-031-55924-2"
}

@article{Constantinou2014,
year = {2014},
month = {apr},
publisher = {},
volume = {506},
number = {1},
pages = {012004},
author = {Constantinou, N C and Lozano-Durán, A and Nikolaidis, M-A and Farrell, B F and Ioannou, P J and Jiménez, J},
title = {Turbulence in the highly restricted dynamics of a closure at second order: comparison with DNS},
journal = {Journal of Physics: Conference Series}
}

@article{Thomas2014,
    author = {Thomas, Vaughan L. and Lieu, Binh K. and Jovanovi\'{c}, Mihailo R. and Farrell, Brian F. and Ioannou, Petros J. and Gayme, Dennice F.},
    title = {Self-sustaining turbulence in a restricted nonlinear model of plane Couette flow},
    journal = {Physics of Fluids},
    volume = {26},
    number = {10},
    pages = {105112},
    year = {2014},
    month = {10},
    issn = {1070-6631},
}

@article{Reynolds1895,
author = {Reynolds, Osborne },
title = {IV. On the dynamical theory of incompressible viscous fluids and the determination of the criterion},
journal = {Phil. Trans. R. Soc. A},
volume = {186},
number = {},
pages = {123-164},
year = {1895}
}

@article{Cavalieri2022,
  title = {Reduced-order Galerkin models of plane Couette flow},
  author = {Cavalieri, Andr\'e V. G. and Nogueira, Petr\^onio A. S.},
  journal = {Phys. Rev. Fluids},
  volume = {7},
  issue = {10},
  pages = {L102601},
  numpages = {10},
  year = {2022},
  month = {Oct},
  publisher = {American Physical Society},
}

\end{document}